%% file: main.tex
\documentclass[fleqn,usenatbib]{mnras}

\usepackage{newtxtext,newtxmath}

\usepackage[T1]{fontenc}

\DeclareRobustCommand{\VAN}[3]{#2}
\let\VANthebibliography\thebibliography
\def\thebibliography{\DeclareRobustCommand{\VAN}[3]{##3}\VANthebibliography}

\usepackage{graphicx}	
\usepackage{amsmath}	
\usepackage{xcolor}     

\title[Nonlinear model of hydrodynamic tori]{Nonlinear dynamics of hydrodynamic tori as a model of oscillations and bending waves in astrophysical discs}

\author[Fairbairn \& Ogilvie]{
Callum W. Fairbairn$^{1}$\thanks{E-mail: cwf29@cam.ac.uk}
and Gordon I. Ogilvie$^{1}$\thanks{E-mail: gio10@cam.ac.uk}
\\
$^{1}$Department of Applied Mathematics and Theoretical Physics, University of Cambridge, Centre for Mathematical Sciences,\\
Wilberforce Road, Cambridge CB3 0WA, UK
}

\date{Accepted 2021 May 25. Received 2021 April 21; in original form 2021 April 21}

\pubyear{2021}

\begin{document}
\label{firstpage}
\pagerange{\pageref{firstpage}--\pageref{lastpage}}
\maketitle

\input{Sections/0_abstract.tex}




\input{Sections/1_introduction}
\input{Sections/2_eulerian}

\input{Sections/3_lagrangian}

\input{Sections/4_linear_modes}

\input{Sections/5_bending_waves}
\input{Sections/6_numerical}

\input{Sections/7_discussion}

\input{Sections/8_conclusion}

\input{Sections/acknowledgements}

\section*{Data Availability}
 
Data used in this paper is available from the authors upon reasonable
request.


\typeout{}
\bibliographystyle{mnras}
\bibliography{main} 


\appendix
\input{Appendices/affine}


\bsp	
\label{lastpage}
\end{document}

%% file: Sections/0_abstract.tex
\begin{abstract}
Understanding oscillations and waves in astrophysical fluid bodies helps to elucidate their observed variability and the underlying physical mechanisms. Indeed, global oscillations and bending modes of accretion discs or tori may be relevant to quasi-periodicity and warped structures around compact objects. While most studies rely on linear theory, observationally significant, nonlinear dynamics is still poorly understood, especially in Keplerian discs for which resonances typically demand a separate treatment. In this work we introduce a novel analytical model which exactly solves the ideal, compressible fluid equations for a non-self-gravitating elliptical cylinder within a local shearing sheet. The aspect ratio of the ring is an adjustable parameter, allowing a continuum of models ranging from a torus of circular cross-section to a thin ring. We restrict attention to flow fields which are a linear function of the coordinates, capturing the lowest order global motions and reducing the dynamics to a set of coupled ordinary differential equations (ODEs). This system acts as a framework for exploring a rich range of hydrodynamic phenomena in both the large amplitude and Keplerian regimes. We demonstrate the connection between tilting tori and warped discs within this model, showing that the linear modes of the ring correspond to oppositely precessing global bending modes. These are further confirmed within a numerical grid based simulation. Crucially, the ODE system developed here allows for a more tractable investigation of nonlinear dynamics. This will be demonstrated in a subsequent paper which evidences mode coupling between warping and vertical motions in thin tilted rings.
\end{abstract}

%% file: Sections/1_introduction.tex
\section{Introduction}
\label{section:introduction}
\subsection{Astrophysical motivation}
Discs appear ubiquitously in the zoo of astrophysical phenomena and occur in a range of systems. From protoplanetary discs to thick accretion flows around black holes, discs are now regarded to be as dynamically important as the central host objects themselves. These structures are inherently dynamic and support a range of oscillations and waves which may be responsible for observational signatures. The key restorative effect which modifies these modes in a non-inertial frame is the Coriolis force, inherited from the rotational motion of the fluid. Whilst a local treatment can categorise a host of wave regimes \citep[e.g.][]{Carroll1985, Kato2001_2}, observationally significant signatures require global coherent modes of large amplitude \citep{Okazaki1987}. 

Several authors have captured such global modes by confining oscillations within tori of limited radial extent. These structures are thought to form around black holes and neutron stars as puffed up hot accretion flows \citep{Frank2002}. The initial theoretical expositions \citep[e.g.][]{Abramowicz1978, Blaes1985} garnered further astrophysical interest in light of the \textit{Rossi X-Ray Timing Explorer} (\textit{RXTE}) observations. Quasi-periodic oscillations (QPOs) in accreting black holes and neutron stars have often been given interpretations that involve oscillating or precessing  rings \citep[e.g.][]{Nowak1997,Stella1999}. High-frequency QPOs sometimes exhibit frequency ratios that have been modelled as the resonances occurring between orbital, vertical and radial oscillations \citep[e.g.][]{Abramowicz2001, Rezzolla2003, Fragile2016, DeAvellar2018}. 

Other oscillating structures have also been identified. In particular warped discs, wherein the orbital plane varies with radius, may explain a range of puzzling observations. Warps can be interpreted as global disc modes which break the mid-plane symmetry of the system and occur whenever a misalignment is present in a system. For example, the seminal work of \cite{Bardeen1975} studied the warped structure due to the misaligned black hole spin and angular momentum axis of the disc. Since then, a range of other scenarios have been investigated. When the magnetic dipole of a central source is tilted with respect to the disc plane it can induce a warp \citep{Lai1999}. Alternatively, companion stars or planets introduce gravitational torques. These may excite tilt in discs via an inclination instability as investigated by \cite{Lubow1992}. This misalignment then allows for the torquing of fluid rings, driving differential precession which manifests as a warp \citep{PapaloizouTerquem1995, Lubow2000}. 

These theoretical efforts have been supported by several key observations. The first indirect observation of a warped disc came from the long period luminosity variations in the Her X-1 X-ray binary source \citep{Katz1973}. The line of sight flux is modulated periodically as light from the source is attenuated by the precessing tilted outer edges. More examples of these so called `superorbital' X-ray binary systems have since been identified by \cite{Kotze2012}. Direct observations were then provided by maser emission which traces the warped structure of discs. The active galaxy NGC4258 (M106) is one such prominent example. This intermediate spiral galaxy exhibits a strong emission feature tracking a warped accretion disc around a central black hole \citep{Miyoshi1995}. 

More recently, interferometric techniques employed by the Atacama Large Millimeter/submillimeter Array (ALMA) have heralded a new era of disc observations. \cite{Sakai2019} reported dust continuum observations of a young protostellar disc with misaligned inner and outer discs, possibly due to anisotropic gas accretion or a misaligned magnetic field. Other protoplanetary discs have proved fertile ground for warp hunting. The \textit{Hubble Space Telescope} images of TW Hya have captured a shadow moving around the outer disc regions at a rate of $22.7 \degr \text{yr}^{-1}$  -- faster than any feature possibly advected with the flow. This suggests an inner tilted precessing disc is blocking light from the central T Tauri star \citep{Debes2017}. ALMA detection of CO molecular line emission also allows for a kinematic study of the gas flow which is effectively modelled with a warped inner structure \citep{Rosenfeld2012}. These warps inferred from shadows are even more exaggerated in transition discs where large radial gaps divide the inner and outer regions and they present extreme inclination differences. The comparison of the HD 142527 gapped protoplanetary disc with parametric radiative transfer models allowed \cite{Marino2015} to deduce a large relative inclination of $70 \degr$ between the inner and outer discs, which may be caused by a companion planet. Similar results have been found for other systems including the DoAr 44 T Tauri transition disc \citep{Casassus2018}. SPHERE+IRDIS observations reveal clear azimuthal dips in reflected infrared polarised intensity which are effectively modelled by an inner disc tilted at $\sim 30 \degr$ . In some systems these distinct rings are thought to form by disc tearing and breaking \citep{Nixon2012}. Indeed, there has been a recent observation of the spectacular triple star system GW Orionis wherein gravitational effects have torn the disc into independently precessing rings \citep{Kraus2020}.

The growing host of observational tilted rings and warps emphasises the need for a comprehensive theoretical understanding of these systems. Whilst there is a body of theory investigating linear warps in discs, the nonlinear dynamics is still poorly understood, particularly in the most important Keplerian regime. The resonances between the vertical and radial motions make this regime trickier to understand but all the more important for the dynamical evolution. 

\subsection{Plan of this paper}

The main purpose of this paper is to introduce a novel ring model which offers a theoretical framework for advancing our understanding of oscillations in tori and warped discs. In particular, this model will allow for nonlinear and resonant phenomena to be investigated, extending the previous body of theory. Whilst thick tori and thin warped disc regimes seem to probe very different physical settings, by varying the aspect ratio of the torus, we expect many of the dynamical results to be smoothly related. 

Here we will focus on the development of the ring model and establish its validity by connecting it with previous linear theory. We begin by localising our fluid equations about a reference circular orbit and then seek axisymmetric, linear flow field solutions which represent the lowest order global modes for an oscillating torus. This reduces to solving a set of coupled ordinary differential equations (ODEs) which exactly describe the flow and shape of the ring. We will also present an alternative formulation of the model using a Lagrangian construction which may prove more amenable to future nonlinear analysis. In order to confirm the correspondence with warped disc theory we will examine the small amplitude modes of the torus and compare these with linear bending waves. These may be understood as global rotating structures exhibiting a hierarchy of precessional timescales. To support these findings, we will briefly present a numerical setup using a grid based solver which is capable of capturing these linear modes and will highlight some of the difficulties in modelling torus oscillations. Finally, we will discuss the connection of our work to previous theoretical efforts. In a subsequent paper, we will apply this model more rigorously to examine the extreme nonlinear dynamics supported by the ring. We find that highly nonlinear mode coupling can occur as energy is interchanged between warping and vertical bouncing motions. These mixed modes merit further attention, in particular with relation to highly warped disc and precessing tilted rings in broken disc systems.

%% file: Sections/2_eulerian.tex
\section{Local Eulerian ring model}
\label{section:eulerian}
\subsection{Localised fluid equations}
We will begin by introducing a fiducial frame which allows for a localisation of the fluid equations. Following the standard shearing box construction \citep[e.g.][]{Hill1878,Hawley1995}, we expand our equations about a local circular reference orbit at $r_0$ with angular velocity $\Omega_0 = \Omega(r_0)$, assuming an axisymmetric potential $\Phi(r,z)$. This orbit has an attached, co-rotating coordinate system $(x,y,z)$ which is defined by $x = (r-r_0)$, $y = r_0(\phi-\Omega_0 t)$ and $z = z$, such that $x$, $y$ and $z$ are the radial, azimuthal and vertical directions respectively. Within the local expansion, for which the size of the domain is much less than $r_0$, the coordinate system is effectively Cartesian and curvature effects are neglected. The equations of motion are then found to be
\begin{gather}
    \label{eulerian_eq:eom_x}
    Du_x-2\Omega_0 u_y=-\partial_x\Phi_\text{t}-\frac{1}{\rho}\partial_x p, \\
   \label{eulerian_eq:eom_y}
    Du_y+2\Omega_0 u_x=-\partial_y\Phi_\text{t}-\frac{1}{\rho}\partial_y p, \\ 
      \label{eulerian_eq:eom_z}
    Du_z=-\partial_z \Phi_\text{t}-\frac{1}{\rho}\partial_z p,  
\end{gather}
where 
\begin{equation}
D = \partial_t + u_x\partial_x + u_y\partial_y + u_z\partial_z
\end{equation}
is the Lagrangian derivative, $\bmath{u}$ is the velocity, $p$ is the pressure and $\rho$ is the density. The local expansion of the tidal potential, $\Phi_\text{t}$, is given by 
\begin{equation}
    \label{eulerian_eq:tidal_potential}
    \Phi_\text{t}=-\Omega_0 S_0 x^2+\frac{1}{2}\nu_0^2 z^2 ,
\end{equation}
where $S_0 = -(r d\Omega/dr)_0$ is the orbital shear rate and $\nu_0^2 = (\partial_{zz} \Phi)_0$ is the vertical oscillation frequency squared of a test particle perturbed from its circular orbit. The first term in $\Phi_\text{t}$ is a reservoir of shear flow energy and describes the destabilising centrifugal contribution arising from the non-inertial acceleration of the orbiting frame. Meanwhile, the second term provides the restorative vertical gravitational term as test particles oscillate harmonically about the mid-plane. A test particle will similarly oscillate radially about the reference orbit at the natural epicyclic frequency $\kappa_0$, thanks to the restorative Coriolis forces appearing in the equation of motion. This is defined by
\begin{equation}
    \label{eulerian_eq:epicylcic_frequency}
    \kappa_0^2 = 2\Omega_0(2\Omega_0-S_0).
\end{equation}
For any spherically symmetric potential, $\nu_0 = \Omega_0$. Furthermore, for a Keplerian disc in a Newtonian point-mass potential, $S_0 = (3/2)\Omega_0$ and so $\kappa_0 = \nu_0 = \Omega_0$. This resonance can lead to distinct dynamical behaviour of interest and should be treated carefully. 

Our model will initially explore the simplest case of an ideal, adiabatic fluid so the dynamical equations are complemented by the mass continuity and energy equations 
\begin{gather}
    \label{eulerian_eq:mass_continuity}
    D\rho = -\rho \Delta, \\
    \label{eulerian_eq:energy}
    Dp = -\gamma p \Delta,
\end{gather}
where
\begin{equation}
    \Delta = \nabla\cdot \bmath{u} 
\end{equation}
is the velocity divergence and $\gamma$ is the adiabatic index (assumed constant). Together these govern the evolution of density and pressure.

\subsection{A simple equilibrium solution}
\label{subsection:eulerian:simple_equilibrium}
We can motivate our general ring solutions by first examining a simple family of equilibrium configurations for the flow. Indeed, the simplest steady state solution of these equations consists of the shear flow $\bmath{u} = -S_0 x\bmath{e_y}$, which describes the circular orbit of fluid particles without any radial pressure gradient support, whilst hydrostatic balance holds in the vertical direction. We can now look for solutions where the shear deviates from this trivial orbital rate and instead consider zonal flows where a \textit{geostrophic} balance holds between the Coriolis force and radial pressure gradients. Let us introduce this zonal flow to be $\bmath{u} = - A x \bmath{e_y}$ where $A$ is some constant defining the linear shear flow. We look for equilibrium solutions in which $\rho$ and $p$ depend only on $(x,z)$. Equations \eqref{eulerian_eq:eom_x}-\eqref{eulerian_eq:eom_z} are then balanced when
\begin{gather}
    \frac{1}{\rho}\partial_x p = 2\Omega_0(S_0-A)x, \\
    \frac{1}{\rho}\partial_z p = -\nu_0^2 z,
\end{gather}
which is equivalent to
\begin{equation}
    \nabla p = -\rho \nabla\psi
\end{equation}
with
\begin{equation}
    \psi = -\Omega_0(S_0-A)x^2+\frac{1}{2}\nu_0^2 z^2.
\end{equation}
It then follows that both $\rho$ and $p$ are functions of $\psi$ only, with the differential relation $dp = -\rho d\psi$. If $A>S_0$, the contours of $\psi$ are similar concentric ellipses and an equilibrium exists in the form of a ring with an elliptical cross-section. This occurs when the shear flow exceeds the orbital shear rate. The aspect ratio of this ring, $\epsilon$, is then defined by the ratio of the radial width to the vertical thickness and is calculated to be
\begin{equation}
    \epsilon = \sqrt{\frac{2\Omega_0(A-S_0)}{\nu_0^2}}.
\end{equation}
Tuning of $A$ then allows for a ring of arbitrary aspect ratio. As the the zonal flow increases, the counteracting pressure gradients must become steeper and the ring becomes more radially confined. Of course, for the choice $A = S_0$, then $\epsilon=0$ and the ring returns to the limiting radially extended case.

\subsection{Dynamical ideal ring solutions}
\label{subsection:eulerian:dynamical_solutions}
These elliptical ring equilibria can be generalised to dynamically evolving tori. We will again look for solutions which are locally axisymmetric and independent of $y$. We assume that the spatial density and pressure can be described by a common function $f(x,z,t)$ which is a material invariant advected by the fluid motion, i.e.
\begin{equation}
    \label{eulerian_eq:material_invariant}
    Df=0.
\end{equation}
Furthermore, to accommodate the changes in density and pressure as the fluid undergoes compressive motions, we write the density and pressure using the separation
\begin{gather}
    \label{eulerian_eq:rho_separation}
    \rho = \hat{\rho}(t)\Tilde{\rho}(f), \\
    \label{eulerian_eq:p_separation}
    p = \hat{p}(t)\Tilde{p}(f).
\end{gather}
We now enforce the key assumption of this framework, positing that the fluid motion is described by a linear flow field in the Cartesian coordinates such that
\begin{equation}
    \label{eulerian_eq:linear_flow}
    u_i = A_{ij}x_j,
\end{equation}
where $A_{ij}$ is a time-dependent, square \textit{flow matrix}. Since the flow is independent of $y$ we also recognise that $A_{i2} = 0$. This allows us to write the divergence as
\begin{equation}
    \label{eulerian_eq:divergence}
    \Delta = A_{11}+A_{33},
\end{equation}
which depends only on time and corresponds to a homogeneous expansion or contraction of the fluid. With these assumptions the mass continuity and energy equations are satisfied provided
\begin{gather}
    \label{eulerian_eq: rho_hat_evolution}
    d_t\hat{\rho} = -\hat{\rho}\Delta, \\
    \label{eulerian_eq: p_hat_evolution}
    d_t\hat{p} = -\gamma \hat{p}\Delta,
\end{gather}
where $d_t = d/dt$. Such a linear velocity field will map ellipses into ellipses. Hence we adopt our materially conserved function $f$ to be a time-dependent quadratic function of the Cartesian coordinates,
\begin{equation}
    \label{eulerian_eq:elliptical_f}
    f = C-\frac{1}{2}S_{ij}x_i x_j,
\end{equation}
where $C$ is some constant and $S_{ij}(t)$ is a time dependent, positive-definite \textit{shape matrix} with $S_{i2}=S_{2i}=0$ in the $y$-independent case. Thus contours of equal density and pressure lie on ellipses akin to the equilibrium discussed in Section \ref{subsection:eulerian:simple_equilibrium}. The shape of these elliptical contours is described by the evolution of $S_{ij}$ which is found by imposing the material conservation condition \eqref{eulerian_eq:material_invariant}. Applying the material derivative to $f$ and using the linear flow field assumption \eqref{eulerian_eq:linear_flow} yields
\begin{equation}
    \left(d_t S_{ij} + S_{ik}A_{kj}+S_{jk}A_{ki}\right)x_ix_j=0.
\end{equation}
This can only be satisfied for all points in space provided
\begin{equation}
    d_t S_{ij} + S_{ik}A_{kj}+S_{jk}A_{ki}=0,
\end{equation}
which reduces to 
\begin{gather}
    \label{equlerian_eq:S11}
    d_t S_{11} + 2(S_{11}A_{11}+S_{13}A_{31})=0, \\
    \label{equlerian_eq:S13}
    d_t S_{13} + S_{11}A_{13}+S_{33}A_{31}+S_{13}\Delta=0,\\
    \label{equlerian_eq:S33}
    d_t S_{33} + 2(S_{13}A_{13}+S_{33}A_{33})=0.
\end{gather}
In order to determine the evolution of the flow matrix, we must return to the equations of motion \eqref{eulerian_eq:eom_x}-\eqref{eulerian_eq:eom_z}. Our assumptions ensure that all the terms are manifestly linear in the Cartesian coordinates, except possibly those involving the pressure gradient which will have the form
\begin{equation}
    -\frac{1}{\rho}\nabla p  = -\frac{\hat{p}}{\hat{\rho}}\frac{1}{\Tilde{\rho}}\frac{d\Tilde{p}}{df}\nabla f.
\end{equation}
Since $\nabla f$ is linear in the coordinates, we can obtain a consistent solution provided $d\Tilde{p}/df\propto\Tilde{\rho}$. Without loss of generality, we can define the functions such that 
\begin{equation}
    \label{eulerian_eq:p_rho_relation}
    \frac{d\Tilde{p}}{df}=\Tilde{\rho},
\end{equation}
whereby the proportionality has been absorbed into the time dependent terms $\hat{\rho}$ and $\hat{p}$. There are many choices available which satisfy this condition and link the density and pressure. An oft used family of solutions is found by assuming a polytropic relationship with index $n$ such that
\begin{equation}
    \label{eulerian_eq: polytropic}
    \Tilde{p}\propto \Tilde{\rho}^{1+1/n}.
\end{equation}
In the case that $n=\infty$ we recover a spatially isothermal ring with a Gaussian distribution of pressure and density. Otherwise, a finite value of $n$ gives a definite ellipse boundary where density and pressure drop to zero and match onto the surrounding vacuum. Indeed, the limiting case $n=0$ describes a homogeneous ring. However, it should be noted that the dynamical evolution of the system doesn't depend on the specific choice of relationship and only requires that equation \eqref{eulerian_eq:p_rho_relation} be satisfied. Using this condition in combination with the previous assumptions, the equation of motion is now satisfied provided
\begin{gather}
    \label{eulerian_eq:A11}
    d_t A_{11}+A_{11}^2+A_{13}A_{31}-2\Omega A_{21} =2\Omega S+\hat{T}S_{11}, \\
    \label{eulerian_eq:A13}
    d_t A_{13}+A_{11}A_{13}+A_{13}A_{33}-2\Omega A_{23}=\hat{T}S_{13}, \\
    \label{eulerian_eq:A21}
    d_t A_{21}+A_{21}A_{11}+A_{23}A_{31}+2\Omega A_{11} =0, \\
    \label{eulerian_eq:A23}
    d_t A_{23}+A_{21}A_{13}+A_{23}A_{33}+2\Omega A_{13} = 0, \\
    \label{eulerian_eq:A31}
    d_t A_{31}+A_{31}A_{11}+A_{33}A_{31}=\hat{T}S_{13}, \\
    \label{eulerian_eq:A33}
    d_t A_{33}+A_{31}A_{13}+A_{33}^2=-\nu^2+\hat{T}S_{33},
\end{gather}
where we have dropped the subscript on the vertical, epicyclic and orbital frequencies and also the orbital shear rate for ease of notation. Here $\hat{T}(t) = \hat{p}/\hat{\rho}$ is a characteristic temperature, which evolves according to 
\begin{equation}
\label{eulerian_eq:t_hat_evolution}
    d_t{\hat{T}} = -(\gamma-1)\hat{T}\Delta,
\end{equation}
which can be shown using equations \eqref{eulerian_eq: rho_hat_evolution} and \eqref{eulerian_eq: p_hat_evolution}.

Together, the evolutionary equations \eqref{equlerian_eq:S11}-\eqref{equlerian_eq:S33} for the shape matrix, \eqref{eulerian_eq:A11}-\eqref{eulerian_eq:A33} for the flow matrix and \eqref{eulerian_eq:t_hat_evolution} for the characteristic temperature, constitute a closed system of 10 first-order, coupled, non-linear ODEs. These govern the dynamics of the ideal ring and despite the various assumptions imposed, we have in fact introduced no approximation. Indeed, these equations present an exact non-linear solution framework to investigate the largest scale global modes of the tori.

%% file: Sections/3_lagrangian.tex
\section{Lagrangian framework}
\label{section:lagrangian}
\subsection{Defining the reference state}
\begin{figure}
    \centering
    \includegraphics[width=\columnwidth]{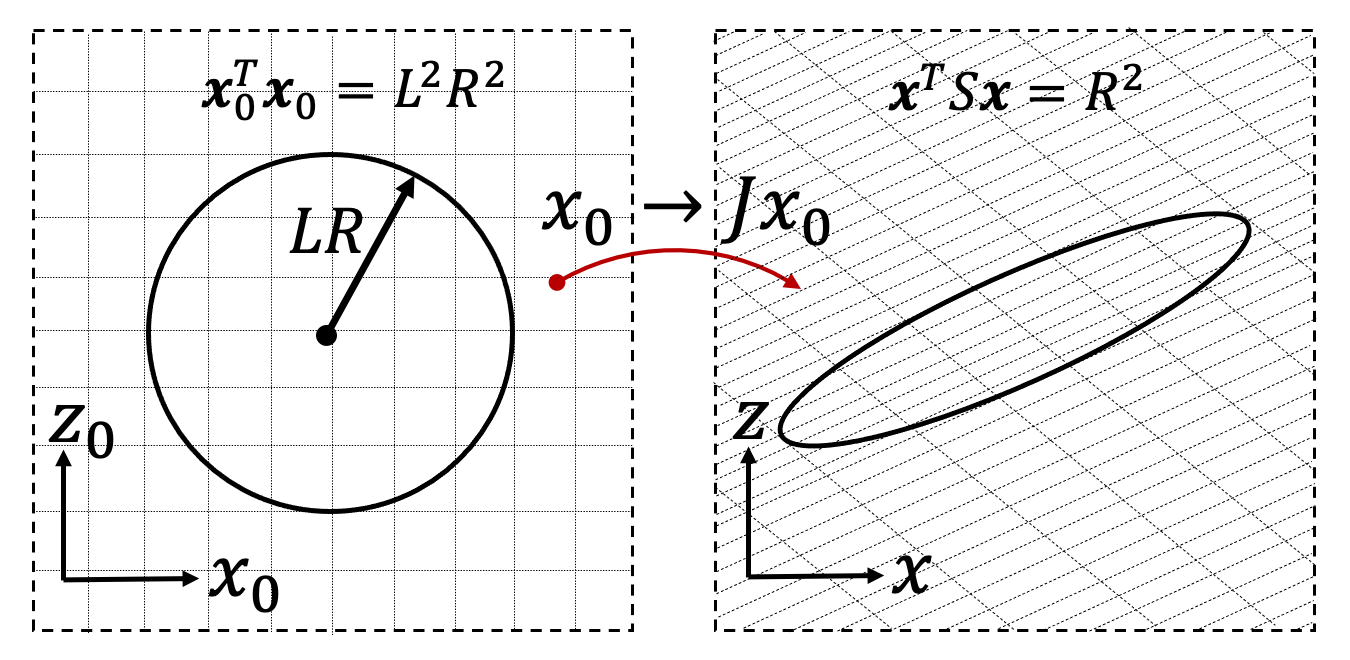}
    \caption{\textit{Left Panel}: This shows the Lagrangian reference state configuration where the solid black line represents a circular isobar. The polar coordinate $R$ parameterises the mass loading onto these contours. The Jacobian matrix enacts an affine transformation to the Eulerian state as shown in the \textit{Right Panel}. The circular contour is stretched as the background grid is deformed.}
    \label{fig:lagrangian_model}
\end{figure}

We can alternatively construct our model from a Lagrangian perspective whereby we track the evolution of individual particles. This will naturally elucidate the conservation properties of the flow and emphasise the underlying oscillatory behaviour of individual fluid particles. Consider a reference state denoted by the Lagrangian coordinates $(x_0,y_0,z_0)$. In general this does not have to be an equilibrium state but is instead an arbitrary, fixed configuration which simply acts to label fluid particles. In order to describe the linear flow fields, we map these fluid particles to their dynamical state by means of a linear transformation denoted
\begin{equation}
    \label{lagrangian_eq:jacobian_transformation}
    \bmath{x}_0 \mapsto \bmath{x}: \quad x_i = J_{ij} x_{0,j},
\end{equation}
where $J_{ij}$ is the time dependent Jacobian matrix. For the assumed axisymmetric setup this will consist of 6 components describing the flow,
\begin{equation}
    \label{lagrangian_eq: jacobian_matrix}
    \mathbfss{J} = 
    \begin{pmatrix}
    J_{11} & 0 & J_{13} \\
    J_{21} & 1 & J_{23} \\
    J_{31} & 0 & J_{33} \\
    \end{pmatrix}.
\end{equation}
Using this formalism, the fluid velocity is simply $\bmath{u} = \dot{\mathbfss{J}}\bmath{x}_0$ so comparison with equation \eqref{eulerian_eq:linear_flow} yields the relation between the flow matrix and the Jacobian elements,
\begin{equation}
    \label{lagrangian_eq:lagrangian_eulerian_flow}
    \mathbfss{A} = \dot{\mathbfss{J}}\mathbfss{J}^{-1}.
\end{equation}
Recall that we require elliptical contours of density, as assumed in the Eulerian model. Thus, without loss of generality, we load mass in the reference state such that particles with constant $f$ lie on circular contours. When these circular fluid rings are acted upon by the linear transformation $\mathbfss{J}$, they will form ellipses with constant $f$. This may be formulated mathematically as
\begin{equation}
    \label{lagrangian_eq:reference_circles}
    \bmath{x}_0^T (\mathbfss{J}^T \mathbfss{S} \mathbfss{J})\bmath{x}_0 = 2(C-f) \equiv R^2,
\end{equation}
with
\begin{equation}
    \label{lagrangian_eq:J_S_relation}
    \mathbfss{J}^T \mathbfss{S} \mathbfss{J} = \frac{\mathbfss{I}}{L^2}.
\end{equation}
Here $\mathbfss{I}$ is the identity matrix and we have introduced $L$ as the characteristic radius of the reference state which is defined by the second mass weighted moment,
\begin{equation}
    L^2 \equiv \langle x_0 x_0 \rangle = \frac{1}{M} \int x_0^2 dm. 
\end{equation}
Here, $M$ is the total mass per unit length and $dm$ denotes integration over the 2D mass distribution. Therefore $R(f)$ defines a dimensionless radius of the circular contours in the reference state. Such a mapping is visualised in Fig.~\ref{fig:lagrangian_model}.
This parameterises the spatial part of the density and pressure separation such that
\begin{equation}
    \label{lagrangian_eq: rho_p_reference_state}
    \rho_0(\bmath{x}_0) = \hat{\rho}_0\Tilde{\rho}(R(\bmath{x}_0)),\quad \text{and} \quad
    p_0(\bmath{x}_0) = \hat{p}_0\Tilde{p}(R(\bmath{x}_0)),
\end{equation}
where $\rho_0$ and $p_0$ denote the density and pressure in the reference state whilst $\hat{\rho}_0$ and $\hat{p}_0$ are characteristic density and pressure factors. Note that we have recast the functional dependence of $\Tilde{\rho}$ and $\Tilde{p}$ from $f$ to $R$ via $f = C-(1/2)R^2$. The action of the Lagrangian mapping then scales area elements by the determinant of the Jacobian matrix,
\begin{equation}
    \label{lagrangian_eq: detJ}
    J\equiv \text{det}(J_{ij}),
\end{equation}
such that the density and pressure transform as
\begin{equation}
    \label{lagrangian_eq:rho_p_transformation}
    \rho = J^{-1} \rho_0, \quad
    p = J^{-\gamma} p_0. 
\end{equation}
Furthermore, the homogeneous compression/expansion of the ring is related to this Jacobian area scaling by 
\begin{equation}
    \label{lagrangian_eq:diveregence}
    \Delta = \frac{d\ln{J}}{dt}.
\end{equation}
This result can be inserted into equation \eqref{eulerian_eq:t_hat_evolution} and integrated to yield the evolution for $\hat{T}$ in terms of the Jacobian elements,
\begin{equation}
    \label{lagrangian_eq:t_hat_integrated}
    \hat{T} = \frac{\hat{T}_0}{J^{\gamma-1}}
\end{equation}
where comparison with equations \eqref{lagrangian_eq: rho_p_reference_state} and \eqref{lagrangian_eq:rho_p_transformation} shows $\hat{T}_0 = \hat{p}_0/\hat{\rho}_0$. Evidently, this integration constant arises as an expression of the conservation of entropy on each fluid particle in the ring.

\subsection{Identifying the Lagrangian}
\label{subsection:identifying_lagrangian}

Understanding the evolution of the ring now reduces to finding the time evolution of the Jacobian transformation elements, $J_{ij}$. We proceed by constructing a Lagrangian which encapsulates the relevant physics. Within the local model this has the form
\begin{equation}
    \label{lagrangian_eq:lagrangian_form}
    \mathcal{L}=\int\left(\frac{1}{2}u^2+2\Omega u_y x-e-\Phi_\text{t}\right)dm
\end{equation}
where the first term represents the kinetic energy, the second term arises from the non-inertial effects of the rotating frame, $e$ is the internal energy and $\Phi_\text{t}$ is the tidal potential. With invariance along $y$, we take $dm$ to represent the mass per unit length and integrate over the total cross-sectional area. Each term must now be reformulated in terms of the Jacobian elements which act as the generalised coordinates of the model. The kinetic energy term is expanded to
\begin{equation}
\label{lagrangian_eq: kinetic}
    \frac{1}{2}\int \left(u_{x}^2+u_{y}^2+u_{z}^2 \right)dm.
\end{equation}
We can express the velocity in terms of the Jacobian elements by switching to the reference state, so for example,
\begin{align}
    \label{lagrangian_eq:lagrangian_kinetic_ux}
    \int u_{x}^2dm  &=\int(\dot{J}_{11}x_0+\dot{J}_{13}z_0)^2 dm \nonumber \\
                    &=M(\dot{J}_{11}^2 \langle x_0x_0 \rangle+2 \dot{J}_{11}\dot{J}_{13} \langle x_0z_0 \rangle +\dot{J}_{13}^2 \langle z_0z_0 \rangle),
\end{align}
where 
\begin{equation}
    \langle x_{0,i} x_{0,j} \rangle = \frac{1}{M}\int x_{0,i} x_{0,j} dm
\end{equation}
is the mass weighted covariance measure of the reference coordinate moments. The circular symmetry of the reference state allows us to identify the second density moments as
\begin{equation}
    \label{lagrangian_eq:lagrangian_covariance}
    \langle x_0 x_0 \rangle = \langle z_0 z_0 \rangle = L^2 ,\quad \langle x_0 z_0 \rangle = 0,
\end{equation}
Inserting this into \eqref{lagrangian_eq:lagrangian_kinetic_ux} and repeating for the remaining terms yields
\begin{equation}
    \label{lagrangian_eq:lagrangian_kinetic}
    \mathcal{L}_{\text{kinetic}}=\frac{ML^2}{2}\left(\dot{J}_{11}^2+\dot{J}_{13}^2+\dot{J}_{21}^2+\dot{J}_{23}^2+\dot{J}_{31}^2+\dot{J}_{33}^2\right).
\end{equation}
One can proceed similarly for the rotational and tidal contributions which respectively give
\begin{gather}
    \label{lagrangian_eq:lagrangian_rotational}
    \mathcal{L}_{\text{rotational}}=2ML^2\Omega\left(J_{11}\dot{J}_{21}+J_{13}\dot{J}_{23}\right), \\
    \label{lagrangian_eq:lagrangian_tidal}
    \mathcal{L}_{\text{tidal}} = ML^2\left[-\Omega S \left(J_{11}^2+J_{13}^2\right)+\frac{1}{2}\nu^2\left(J_{31}^2+J_{33}^2\right)\right].
\end{gather}
Finally we examine the internal energy contribution. Using the ideal equation of state,
\begin{equation}
    \label{lagrangian_eq:ideal_eos}
    e = \frac{p}{\rho(\gamma-1)},
\end{equation}
the energy density integral becomes
\begin{equation}
    \mathcal{L}_{e}=\int e dm = \frac{1}{(\gamma-1)}\int p\,dx\,dz.
\end{equation}
Inserting the separation ansatz \eqref{eulerian_eq:p_separation} and transforming to Lagrangian coordinates gives
\begin{equation}
    \mathcal{L}_{e} = \frac{\hat{p} J}{\gamma-1}\int\Tilde{p} dx_0 dz_0,
\end{equation}
where the area scaling incurs a factor of $J$. In order to progress we recall that the spatial pressure term is constant on circular Lagrangian contours as per equation \eqref{lagrangian_eq: rho_p_reference_state}, so $\Tilde{p}=\Tilde{p}(R)$. Thus it is convenient to switch to polar Lagrangian coordinates such that
\begin{equation}
    x_0^2+z_0^2=L^2R^2, \quad
    (x_0,z_0) = LR(\cos{\theta},\sin{\theta}),
\end{equation}
where $\theta$ is the usual polar angle in the Lagrangian reference state and so
\begin{equation}
    \label{lagrangian_eq:lagrangian_e_intermediate_step}
    \mathcal{L}_{e}=\frac{\hat{p} J L^2}{\gamma-1}\int \Tilde{p}(R) R dR d\theta.
\end{equation}
Using condition \eqref{eulerian_eq:p_rho_relation} allows us to relate $\Tilde{p}$ and $\Tilde{\rho}$ via
\begin{equation}
    \label{lagrangian_eq: p_rho_R_relation}
    \frac{d\Tilde{p}}{dR}=-R\Tilde{\rho}.
\end{equation}
Integrating equation \eqref{lagrangian_eq:lagrangian_e_intermediate_step} by parts, inserting the expression above and converting back to an integral over mass gives
\begin{equation*}
     \mathcal{L}_{e} = \frac{\hat{p} J}{(\gamma-1)}\int\frac{L^2 R^2}{2}\Tilde{\rho} R dR d\theta
     = \frac{\hat{p}}{\hat{\rho}(\gamma-1)L^2}\int\frac{x_0^2+z_0^2}{2} dm.
\end{equation*}
The integral can now be computed from the standard covariance results \eqref{lagrangian_eq:lagrangian_covariance} and the ratio $\hat{p}/\hat{\rho}$ is given by equation \eqref{lagrangian_eq:t_hat_integrated} such that
\begin{equation}
    \label{lagrangian_eq:lagrangian_internal}
    \mathcal{L}_{e} = \frac{M \hat{T}_0}{(\gamma-1)J^{\gamma-1}}.
\end{equation}
Finally, summing all contributions \eqref{lagrangian_eq:lagrangian_kinetic}, \eqref{lagrangian_eq:lagrangian_rotational}, \eqref{lagrangian_eq:lagrangian_tidal} and \eqref{lagrangian_eq:lagrangian_internal} and cancelling the $ML^2$ factor gives the total Lagrangian,
\begin{align}
    \label{lagrangian_eq:total_lagrangian}
    \mathcal{L} &= \frac{1}{2}(\dot{J}_{11}^2+\dot{J}_{13}^2+\dot{J}_{21}^2+\dot{J}_{23}^2+\dot{J}_{31}^2+\dot{J}_{33}^2)-\frac{\hat{T}_0}{(\gamma-1) J^{\gamma-1}L^2} \nonumber \\
    &+\Omega S (J_{11}^2+J_{13}^2)-\frac{1}{2}\nu^2(J_{31}^2+J_{33}^2)+2\Omega(J_{11}\dot{J}_{21}+J_{13}\dot{J}_{23}).
\end{align}
We can now derive the dynamical evolution by means of the usual Euler-Lagrange equations
\begin{equation}
    \label{lagrangian_eq:euler_lagrange_equations}
    \frac{d}{dt}\left(\frac{\partial \mathcal{L}}{\partial \dot{J}_{ij}}\right) = \frac{\partial \mathcal{L}}{\partial J_{ij}}.
\end{equation}
This gives
\begin{align}
    \label{lagrangian_eq:J11_ode}
    & \ddot{J}_{11} = 2\Omega\dot{J}_{21}+2\Omega S J_{11}+\frac{\hat{T}_0}{J^\gamma L^2}J_{33}, \\
    \label{lagrangian_eq:J13_ode}
    & \ddot{J}_{13} = 2\Omega\dot{J}_{23}+2\Omega S J_{13}-\frac{\hat{T}_0}{J^\gamma L^2}J_{31}, \\
    \label{lagrangian_eq:J21_ode}
    & \ddot{J}_{21} = -2\Omega\dot{J}_{11}, \\
    \label{lagrangian_eq:J23_ode}
    & \ddot{J}_{23} = -2\Omega\dot{J}_{13}, \\
    \label{lagrangian_eq:J31_ode}
    & \ddot{J}_{31} = -\nu^{2}J_{31}-\frac{\hat{T}_0}{J^\gamma L^2}J_{13}, \\
    \label{lagrangian_eq:J33_ode}
    & \ddot{J}_{33} = -\nu^{2}J_{33}+\frac{\hat{T}_0}{J^\gamma L^2}J_{11}.
\end{align}
Formally this is a coupled system of 6 non-linear, second order differential equations which govern the evolution of the flow. These can in fact also be derived as a specific sub-case of the general Lagrangian framework for the affine motion of discs as developed by \cite{Ogilvie2018} and discussed in appendix \ref{appendices:affine}.

\subsection{Conserved circulation integrability}

Equations \eqref{lagrangian_eq:J21_ode} and \eqref{lagrangian_eq:J23_ode} are clearly integrable, allowing for the reduction in order of the system of equations. This naturally arises as a consequence of conservation laws which restrict the degrees of freedom of this system. The conservation of entropy has already introduced the constant $\hat{T}_0$ into equation \eqref{lagrangian_eq:t_hat_integrated}. Now, we will make use of the conservation of circulation to interpret the integrability of the $J_{21}$ and $J_{23}$ terms. The Poincar\'e-Bjerknes circulation theorem extends Kelvin's result to the rotating shearing box reference frame such that
\begin{equation}
    \label{lagrangian_eq:circulation_theorem}
    \frac{D\Gamma(t)}{Dt} = 0,
\end{equation}
where
\begin{equation}
    \label{lagrangian_eq:circulation}
    \Gamma = \oint_{\mathcal{C}(t)} \left( \mathbf{u}+\mathbf{\Omega}\times \mathbf{x} \right)\cdot d\mathbf{x},
\end{equation}
with the line integral taken around the material loop $\mathcal{C}(t)$. Since material contours are fixed in the Lagrangian reference frame it is convenient to switch to this perspective. Using the transformation $u_i = \dot{J}_{ij} x_{0,j}$ we have that
\begin{equation}
    \label{lagrangian_eq:circulation_jacobian}
    \Gamma = (\dot{J}_{ij} J_{ik} + \Omega(J_{1j}J_{2k}-J_{1k}J_{2j}))\oint_{\mathcal{C}_0} x_{0,j}dx_{0,k}.
\end{equation}
For any material contour $\mathcal{C}_0$, the integral can be evaluated to give a constant matrix 
\begin{equation}
    I_{jk} = \oint_{\mathcal{C}_0} x_{0,j}dx_{0,k},
\end{equation}
which encapsulates the relation between the Jacobian coordinates through \eqref{lagrangian_eq:circulation_jacobian}. This matrix is evaluated using Stokes' Theorem which gives
\begin{equation}
    I_{jk} = \epsilon_{ijk} A_{0,i},
\end{equation}
where, $\epsilon_{ijk}$ is the anti-symmetric Levi-Civita symbol and $\mathbfit{A}_0$ is the vector area of any open surface bounded by $\mathcal{C}_0$. We are free to choose the material contours however we like, so in particular we may take $A_{0,1} \equiv A_{x}$ and $A_{0,2} = A_{0,3} = 0$ so that the contour lies entirely in the $y-z$ plane and has a conserved circulation denoted $\Gamma_x$. Inserting this into \eqref{lagrangian_eq:circulation_jacobian} gives
\begin{equation}
    \dot{J}_{23} = -\frac{\Gamma_x}{A_x} - 2\Omega J_{13},
\end{equation}
where $-\Gamma_x/A_x \equiv C_x$ is obviously the constant identified by integrating \eqref{lagrangian_eq:J23_ode}. Equally we can choose $A_{0,3} \equiv A_{z}$ and $A_{0,1} = A_{0,2} = 0$ so the contour lies entirely in the $x-y$ plane and the conserved circulation is denoted $\Gamma_z$. In this case
\begin{equation}
    \dot{J}_{21} = \frac{\Gamma_z}{A_z} - 2\Omega J_{11},
\end{equation}
and $\Gamma_z/A_z \equiv C_z$ is the constant arising from the integration of \eqref{lagrangian_eq:J21_ode}. Using this integrability, we may reduce equations \eqref{lagrangian_eq:J11_ode} -- \eqref{lagrangian_eq:J33_ode} to 
\begin{align}
   & \ddot{J}_{11} +\kappa^2 J_{11}=2C_z+\frac{\hat{T}_0}{J^\gamma L^2}J_{33}, \label{nonlinear_eq: J_11_ode}\\
   & \ddot{J}_{13} +\kappa^2 J_{13}=2C_x-\frac{\hat{T}_0}{J^\gamma L^2}J_{31}, \label{nonlinear_eq: J_13_ode}\\
   & \ddot{J}_{31} +\nu^2 J_{31}=-\frac{\hat{T}_0}{J^\gamma L^2}J_{13}, \label{nonlinear_eq: J_31_ode}\\
   & \ddot{J}_{33} +\nu^2 J_{33}=\frac{\hat{T}_0}{J^\gamma L^2}J_{11}, \label{nonlinear_eq: J_33_ode}
\end{align}
where the shear rate has been eliminated in favour of the epicylic frequency. These equations clearly highlight the harmonic oscillatory structure inherited from the simple test particle motion. These are then coupled together by means of the pressure term on the right hand side.

%% file: Sections/4_linear_modes.tex
\section{Linear modes}
\label{section:linear}

Having developed the ring model, we may now examine the simplest linear dynamics. To this end we will explore the linear modes supported by the system. Note that the linear transformations permitted by equation \eqref{eulerian_eq:linear_flow} leave the centre of mass of the ring fixed at the origin of the shearing box. Thus we are neglecting transformations which translate the whole ring vertically or radially. Such vertical oscillations at the vertical frequency, $\nu$, would simply correspond to a global inclination of the ring with respect to the chosen shearing box orbital plane. Meanwhile, radial oscillations at the epicylic frequency would describe an eccentric offset of the torus. Here we focus instead on the physically interesting breathing modes which engage the thermodynamics of the ring and tilting modes which rock back and forth about the centre of the shearing box.

\subsection{Eulerian perspective}

In accordance with Section \ref{subsection:eulerian:simple_equilibrium} we can identify the equilibrium solutions about which to perform the perturbation analysis. We take $S_{ij}$ to be constant with $S_{13} = 0$, such that the ring has an aspect ratio $\epsilon = \sqrt{S_{11}/S_{33}}$. This shape is supported by a geostrophic balance where the flow matrix vanishes apart from the $A_{21}$ component. In this case equilibrium is achieved provided
\begin{equation}
    \label{linear_eq:eulerian_equilibrium}
    -2\Omega(S+A_{21}) = \hat{T}S_{11} \quad \text{and} \quad \nu^2 =\hat{T}S_{33}.    
\end{equation}
We can now introduce small perturbations about this equilibrium ring in the form $A_{ij} = A_{e,ij}+ A_{ij}'$, $S_{ij} = S_{e,ij}+S_{ij}'$ and $\hat{T} = \hat{T}_{e}+\hat{T}'$, where the subscript $e$ denotes the equilibrium background quantity. Inserting into \eqref{equlerian_eq:S11} -- \eqref{equlerian_eq:S33} and  \eqref{eulerian_eq:A11}--\eqref{eulerian_eq:t_hat_evolution} and dropping non-linear terms yields the matrix equation
\begin{equation}
    \frac{d\mathbf{X}}{dt} = M\mathbf{X}, 
\end{equation}
where
\begin{equation}
    \mathbf{X}^T = [S_{11}', S_{13}', S_{33}', A_{11}', A_{13}', A_{21}', A_{23}', A_{31}', A_{33}', \hat{T}'],
\end{equation}
and $M$ is a sparse $10\times10$ matrix which depends on the background equilibrium. We look for oscillatory solutions with time dependence $\exp [i\omega t]$, which results in the eigenvalue problem $M\mathbf{X} = i\omega \mathbf{X}$. Solving this gives a 10th order dispersion relation for the 10 associated eigenmodes. Alternatively a more enlightening approach is to identify the separate mode families by carefully motivating the form of the initial perturbations.

\subsubsection{Breathing modes}

Restricting our attention to breathing modes which preserve the $x$ and $z$ reflectional symmetries of the ring, we set $S_{13}' = A_{13}'= A_{31}' = A_{23}' = 0$. This reduces the eigenvalue problem to a 6th order system. Taking suitable combinations of these equations allows further simplification as we are able to eliminate $S_{11}'$, $S_{33}'$ and $\hat{T}'$ in favour of the flow variables $A_{11}'$ and $A_{33}'$. Meanwhile, $A_{21}'$ simply responds to $A_{11}'$ as radial flows drive azimuthal perturbations, arising as a consequence of angular momentum conservation. This yields the simple eigenvalue problem
\begin{equation}
    \omega^2
    \begin{pmatrix}
    A_{11}' \\
    A_{33}'
    \end{pmatrix} = 
    \begin{pmatrix}
    \kappa^2+\gamma \epsilon^2\nu^2 & \nu^2(\gamma-1)\epsilon^2 \\
    (\gamma-1)\nu^2 & (\gamma+1)\nu^2
    \end{pmatrix}
    \begin{pmatrix}
    A_{11}' \\
    A_{33}'
    \end{pmatrix}.
\end{equation}
The associated 4th order characteristic equation gives a pair of frequencies
\begin{multline}
    \label{linear_eq:breathing_modes_dispersion}
    \omega^2 = \frac{1}{2}\Big[\Big.\kappa^2+(1+\gamma+\gamma\epsilon^2)\nu^2 \\ \pm\sqrt{(\kappa^2+(1+\gamma+\gamma\epsilon^2)\nu^2)^2-4((1+\gamma)\kappa^2\nu^2+(3\gamma-1)\epsilon^2\nu^4)}\Big.\Big].
\end{multline}
These modes generally couple the vertical and radial directions as indicated by the presence of both natural frequencies $\kappa$ and $\nu$. Furthermore, the dependence on the adiabatic index signifies that the modes are compressive in nature. Indeed, as we let $\gamma \rightarrow \infty$ in the incompressible limit, the frequencies become infinite as the sound speed diverges. In the opposite isothermal regime, $\gamma=1$ and the radial and vertical modes decouple. The radial modes take on the form of an inertial-acoustic oscillation whilst the vertical mode reduces to the usual $\omega=\sqrt{2}\nu$ as expected for perturbed isothermal discs \citep{Ogilvie2013}. We also note the dependence on the aspect ratio of the ring. As $\epsilon\rightarrow 0$ the ring becomes extended and the radial pressure gradients are negligible. Thus the radial oscillation breathing mode is no longer driven by the vertical motion and the frequency reduces to $\kappa$. 

\subsubsection{Zero frequency modes}

The above analysis identifies 4 eigenmodes out of the 6 available for the 6th order system. The remaining 2 modes correspond to the zero frequency perturbations which essentially shift the initial equilibrium to a neighbouring equilibrium ring. The fact these zero frequency modes appear within the breathing mode analysis is to be expected since the equilibrium configurations all possess even $x-z$ parity so are reached by symmetric perturbations. This shift to a neighbouring equilibrium is constrained by the conservation laws which are associated with each zero frequency mode. The conservation of entropy ensures that
\begin{equation}
    \label{linear_eq:entropy_conservation}
    \frac{d}{dt}\left(\frac{1}{2}\frac{S_{11}'}{S_{e,11}}+\frac{1}{2}\frac{S_{33}'}{S_{e,33}}-\frac{1}{\gamma-1}\frac{\hat{T}'}{\hat{T}}\right) = 0,
\end{equation}
whilst the conservation of angular momentum enforces
\begin{equation}
    \label{linear_eq:momentum_conservation}
    \frac{d}{dt}\left( \frac{A_{21}'}{2\Omega+A_{e,21}}-\frac{S_{11}'}{2S_{11}}\right) = 0.
\end{equation}

\subsubsection{Tilting modes}
\label{subsubsection:tilting_modes}

\begin{figure}
    \centering
    \includegraphics[width=\columnwidth]{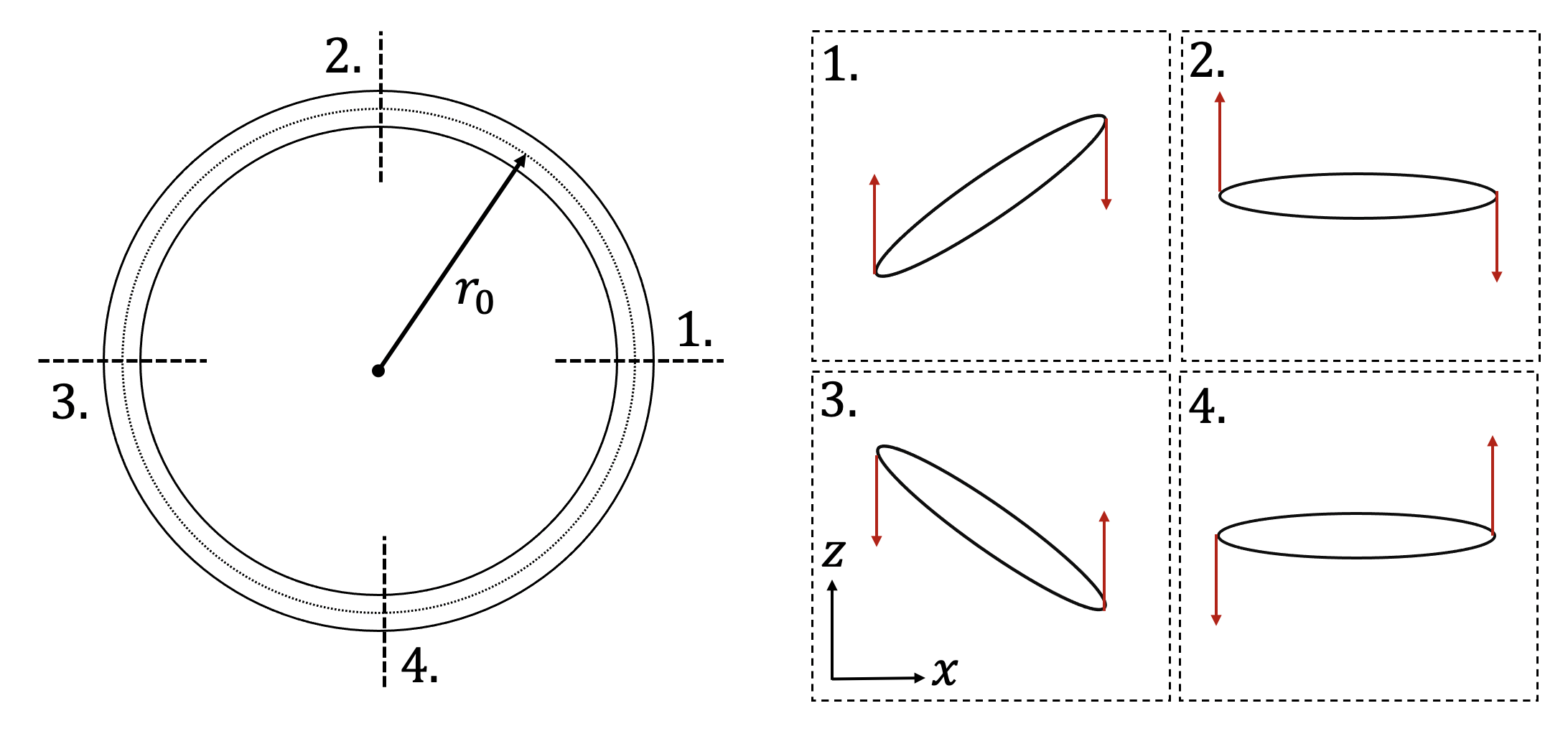}
    \caption{\textit{Left panel}: Shows a top down cartoon of the slender ring in the global orbital plane. The numbered quadrants denote different locations around the anti-clockwise orbit. \textit{Right panel}: Visualises the local model expanded about $r_0$, as seen in the $x-z$ slices cutting through the ring in each quadrant location. The red arrows show the future evolution of the tilting ellipse as locally seen by an observer orbiting around the fixed global structure. Connecting these slices shows how a global warp is equivalent to a local tilting mode.}
    \label{fig:tilting_modes}
\end{figure}

The remaining 4 modes associated with the full 10th order system are now identified by introducing perturbations which break the reflectional symmetry. Retaining the off-diagonal terms and setting $A_{11}'= A_{33}'=A_{21}'=S_{11}'=S_{33}'=\hat{T}' = 0$ yields a 4th order eigenvalue problem for the remaining variables. Eliminating $A_{23}'$ and $S_{13}'$ in favour of $A_{13}'$ and $A_{31}'$ gives the eigenvalue problem as
\begin{equation}
    \omega^2
    \begin{pmatrix}
    A_{13}' \\
    A_{31}'
    \end{pmatrix} = 
    \begin{pmatrix}
    \kappa^2 & \nu^2 \\
    \nu^2 \epsilon^2 & \nu^2
    \end{pmatrix}
    \begin{pmatrix}
    A_{13}' \\
    A_{31}'
    \end{pmatrix}.
\end{equation}
The associated frequencies are then given by
\begin{equation}
    \label{linear_eq:tilting_modes_dispersion}
    \omega^2=\frac{1}{2}\left(\kappa^2+\nu^2\pm\sqrt{\left(\kappa^2-\nu^2\right)^2+4\epsilon^2\nu^4}\right).
\end{equation}
These tilting modes break the symmetry of the ring as the major and minor axes are now inclined with respect to the mid-plane. The appearance of $\kappa$ and $\nu$ again signifies the strong coupling of the vertical and radial oscillations. However, now the absence of $\gamma$ implies that they are incompressible. This is indeed the case as $A_{11} = A_{33}=0$ ensures that the divergence $\Delta = 0$, in accordance with equation \eqref{eulerian_eq:divergence}. For small aspect ratios the approximate solutions are
\begin{equation}
    \label{linear_eq:tilting_modes_non_resonant}
    \omega^2\approx\kappa^2+\epsilon^2\frac{\nu^4}{\kappa^2-\nu^2}, \quad 
    \omega^2\approx\nu^2+\epsilon^2\frac{\nu^4}{\nu^2-\kappa^2},
\end{equation}
which shows a distinction between the dominant radial epicyclic shearing mode of the ring as opposed to the dominant vertical tilting mode of the ring. However, in the Keplerian case $\kappa=\nu$ and there is resonant degeneracy between the epicyclic frequencies so instead
\begin{equation}
    \label{linear_eq:tilting_modes_resonant}
    \omega^2 = (1\pm\epsilon)\nu^2.
\end{equation}
When viewed from the local model, these modes appear to rock back and forth periodically about the centre of the shearing box. However an alternative perspective is to Doppler shift these solutions back into the global reference frame. Now the dynamical tilting can be interpreted as the passage of the global ring structure seen by an observer moving around on the fiducial orbit. For example, a stationary global torus, with some azimuthally dependent tilt about this orbit, will appear to oscillate within the local model. This is visualised in figure \ref{fig:tilting_modes}. It is worth re-emphasising the key but subtle distinction between a local tilt of the elliptical cross-section about the centre of the shearing box, versus a global inclination of the whole ring structure which is simply a trivial rotation within a spherically symmetric potential. In this paper a tilting motion refers to the former, non-trivial mode and is viewed as a warping of the disc about the shearing box orbital plane, as will be explored further in section \ref{section:bending_waves}. 

%% file: Sections/5_bending_waves.tex
\section{Connection with linear bending waves}
\label{section:bending_waves}
\subsection{Bending wave theory}
The linear tilting modes identified in Section \ref{subsubsection:tilting_modes} are analogous to warped structures in discs. Indeed, warps have previously been examined within the context of a local model by \cite{Ogilvie2013}. They note that a fixed, global warp appears as a uniform rocking of the disc within the local model. Similarly in this ring model, the tilting torus will further illuminate the behaviour of warps. To this end we will formally illustrate the correspondence between our tilting modes and linear warps, as described by the bending wave theory of \cite{Lubow2000}. Let us restrict our attention to the thin disc, inviscid regime without nodal precession such that $\nu = \Omega$.  Within this framework, the warp structure in the global inertial frame is governed by a pair of partial differential equations,
\begin{equation}
    \Sigma r^2\Omega \frac{\partial \mathbfit{l}}{\partial t} = \frac{1}{r}\frac{\partial \mathbfit{G}}{\partial r},
\end{equation}
\begin{equation}
    \frac{\partial\mathbfit{G}}{\partial t} - \omega_a \mathbfit{e}_z\times\mathbfit{G} = \frac{P r^3\Omega}{4}\frac{\partial \mathbfit{l}}{\partial r},
\end{equation}
where $\Sigma$ and $P$ are the vertically integrated density and pressure respectively. $\mathbfit{l}$ and $\mathbfit{G}$ are then the horizontal components of the unit tilt vector and the internal torque, which describe the evolution of the warp. The term $\omega_a \equiv (\Omega^2-\kappa^2)/2\Omega$ encapsulates the detuning from Keplerian resonance which results in an apsidal precession. Indeed, when the epicyclic frequency only differs slightly from the orbital frequency, $\kappa\approx\Omega-\omega_a$ to leading order. Now, in order to compare with our local model we must expand these equations about the reference radius $r_0$:
\begin{equation}
    \Sigma(x) r_0^2\Omega_0\frac{\partial \mathbfit{l}}{\partial t} = \frac{1}{r_0}\frac{\partial \mathbfit{G}}{\partial x},
\end{equation}
\begin{equation}
    \frac{\partial\mathbfit{G}}{\partial t} - \omega_a \mathbfit{e}_z\times\mathbfit{G} = \frac{P(x) r_0^3\Omega_0}{4}\frac{\partial \mathbfit{l}}{\partial x}.
\end{equation}
These equations can be combined into a single, second order PDE for the tilt vector. This has two independent components so is amenable to a complex description,
\begin{equation}
    \label{bending_eq:complex_tilt_evolution}
    \Sigma(x)\frac{\partial^2 W}{\partial t^2}-\frac{1}{4}\left( \frac{\partial P}{\partial x}\frac{\partial W}{\partial x}+P\frac{\partial^2 W}{\partial x^2}\right) - i \Sigma(x)\omega_a \frac{\partial W}{\partial t} = 0,
\end{equation}
where $W = l_x+i l_y$ is the complex tilt variable. The vertically integrated density and pressure structure is now set in correspondence with the ring model. We perform a separation of variables as before such that
\begin{equation}
    \label{bending_eq: P_separation}
    P = \hat{p}\Tilde{P} \quad \text{where} \quad \Tilde{P} = \int\Tilde{p} dz
\end{equation}
and
\begin{equation}
    \Sigma = \hat{\rho}\Tilde{\Sigma} \quad \text{where} \quad \Tilde{\Sigma} = \int\Tilde{\rho} dz
\end{equation}
Note once again the ratio $\hat{p}/\hat{\rho}=\hat{T}$. Differentiating \eqref{bending_eq: P_separation},
inserting \eqref{eulerian_eq:p_rho_relation} and using the definition of $f$  given by equation \eqref{eulerian_eq:elliptical_f} yields
\begin{equation}
    \frac{\partial P}{\partial x} = -S_{11}x\hat{p}\Tilde{\Sigma} = -S_{11}x\hat{T}\Sigma.
\end{equation}
Now, recall equation \eqref{linear_eq:eulerian_equilibrium} for the hydrostatic equilibrium $S_{33}\hat{T} = \nu^2 = \Omega^2$. Thus, $S_{11}\hat{T} = \epsilon^2\Omega^2$. Using this result and inserting into equation \eqref{bending_eq:complex_tilt_evolution} leads to 
\begin{equation}
    \label{bending_eq:}
    \frac{\partial^2 W}{\partial t^2}+\frac{1}{4}\epsilon^2\Omega^2\frac{\partial W}{\partial x}x-\frac{1}{4}\frac{P}{\Sigma}\frac{\partial^2 W}{\partial x^2}-i\omega_a\frac{\partial W}{\partial t} = 0.
\end{equation}
We now seek oscillatory tilting solutions which are linear in the coordinate $x$. These will take the form $W = W_0\exp[i\omega_p t]x$. Inserting this ansatz yields the frequencies, 
\begin{equation}
    \label{bending_eq:dispersion}
    \omega_p = \frac{\omega_a}{2}\pm\frac{1}{2}\sqrt{\omega_a^2+\epsilon^2\Omega^2} .
\end{equation}
Physically these oscillations correspond to a rotation of the tilt vector according to
\begin{equation}
    \label{bending_eq:rotating_tilt_vector}
    \omega_p t = \arctan\left(\frac{l_y}{l_x}\right)+\phi_0,
\end{equation}
where $\phi_0$ is some arbitrary phase offset. Thus we interpret $\omega_p$ as a precessional frequency - as the warped global structure slowly rotates, an observer fixed at a point in the ring will see it rock back and forth. Clearly $\omega_p>0$ and $\omega_p<0$ result in prograde and retrograde precession respectively. In the limit $\epsilon\rightarrow 0$ the ring reduces to a thin disc with $\omega_p=0$ or $\omega_p = \omega_a$. Thus it permits a stationary warped structure or a shearing mode which is driven by the radial apsidal precession. In the Keplerian resonant regime with $\omega_a = 0$ we see that $\omega_p = \pm\epsilon\Omega/2$ so the precession is driven entirely by the radial pressure gradients set up due to the finite extent of the ring. 

\subsection{Identification with tilting ring modes}

Indeed we can formally identify these bending modes with the linear tilt solutions identified in Section \ref{subsubsection:tilting_modes}. The bending wave analysis is performed about some local radius in a non-rotating, inertial frame where the warped structure corresponds to a global pattern with azimuthal wavenumber $m=1$. In order to compare with the frequency observed in the orbiting frame, we must perform a Doppler shift which changes the frequency by 1 times $\Omega$ such that
\begin{equation}
    \label{bending_eq:dopppler_shift}
    \omega' \equiv \omega_p-\Omega.
\end{equation}
The linear bending wave theory is formally valid in the case of thin discs with small deviations from Keplerian resonance,
\begin{equation}
    \kappa^2 = \Omega^2(1+\delta) \implies \omega_a = -\frac{\delta\Omega}{2}
\end{equation}
where $\delta\ll 1$ parameterises the deviation from resonance. Thus $\epsilon$ and $\omega_a$ are both small. We now focus on two regimes in turn. First consider the non-resonant regime for which $\epsilon/\delta \ll 1$. In this case
\begin{equation}
    \omega_p \approx \frac{1}{2}\omega_a\left[ 1\pm\left( 1+\frac{\epsilon^2\Omega^2}{2\omega_a^2}\right) \right].
\end{equation}
Taking the ($+$) prograde solution and Doppler shifting yields
\begin{align}
|\omega'| &\approx \Omega-\omega_a-\frac{\epsilon^2\Omega^2}{4\omega_a^2} \approx \kappa+\frac{\epsilon^2\Omega^2}{2(\kappa^2-\Omega^2)}+O(\delta^2).
\end{align}
This will match onto the result found for the non-resonant tilting mode given by equation \eqref{linear_eq:tilting_modes_non_resonant} provided the $O(\delta^2)$ terms are negligible. Taking $\delta\sim\epsilon^X$ and balancing the order of the last two terms yields the range of agreement $2/3<X<1$. Meanwhile the (-) retrograde mode gives
\begin{align}
|\omega'| &\approx \Omega+\frac{\epsilon^2\Omega^2}{2(\Omega^2-\kappa^2)}.
\end{align}
These match onto the results found for the non-resonant tilting modes in \eqref{linear_eq:tilting_modes_non_resonant}. Similarly, we can examine the resonant case where $\omega_a = 0$ and $\omega_p = \pm\epsilon\Omega/2$. The Doppler shifted frequency is then
\begin{equation}
    |\omega'| = \left(1\pm\frac{\epsilon}{2}\right)\Omega,
\end{equation}
which agrees with the resonant ring tilting modes found in \eqref{linear_eq:tilting_modes_resonant}. 
\begin{figure}
    \centering
    \includegraphics[width=\columnwidth]{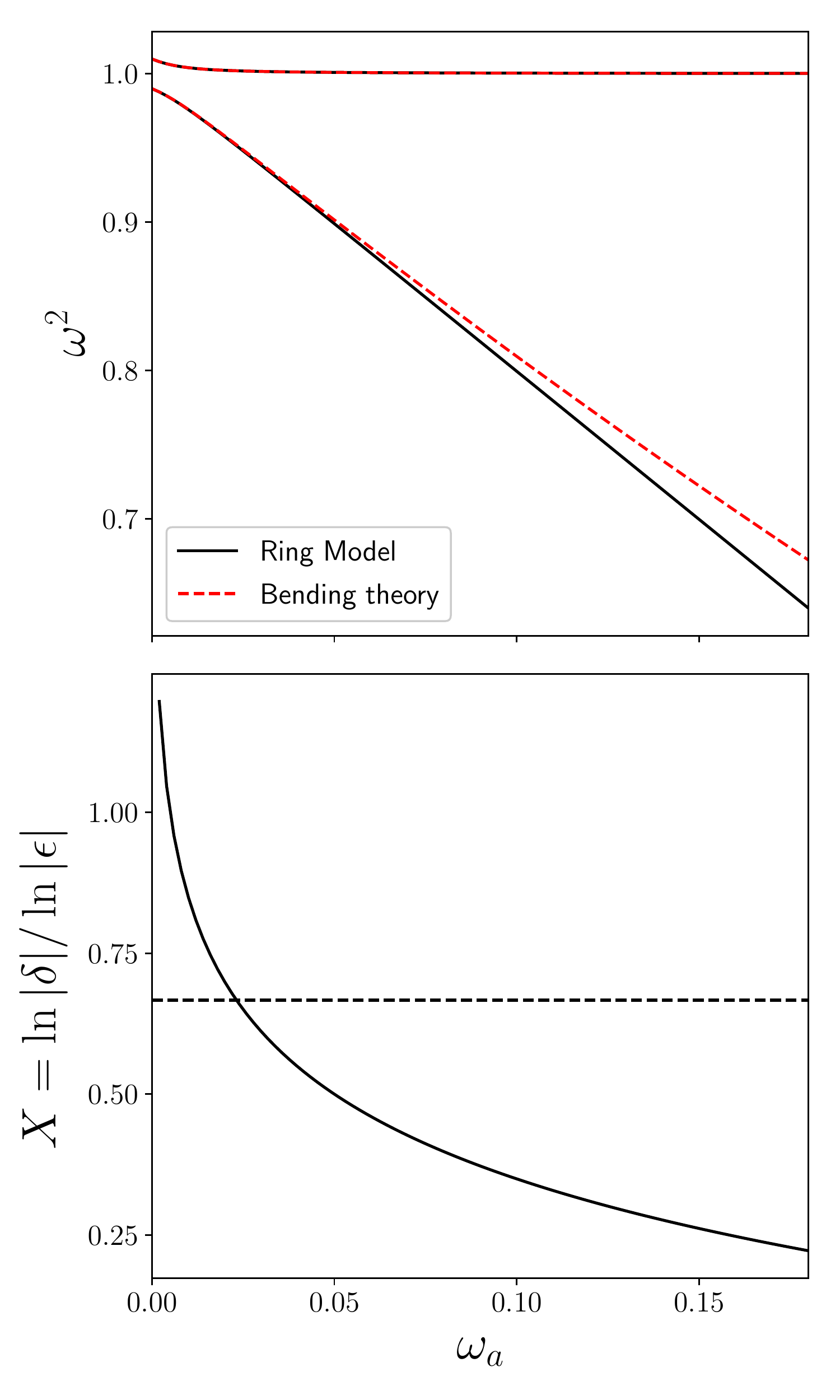}
    \caption{The \textit{upper panel} shows how the linear frequency $\omega^2$ varies with apsidal precession $\omega_a$ for a ring with aspect ratio $\epsilon=0.01$. The black line plots $\omega^2$ as calculated from the linear frequency relation found for our ring model in equation \eqref{linear_eq:tilting_modes_dispersion}. Meanwhile the red dashed line shows the Doppler shifted linear bending mode frequency $\omega'^2$, found by combining equations \eqref{bending_eq:dispersion} and \eqref{bending_eq:dopppler_shift}. The lower panel computes the variation in the asymptotic scaling parameter $X$. The dashed line denotes where this drops below the $2/3$ threshold, at which point the lower branch non-resonant solutions differ between the ring model and bending wave theory.}
    \label{fig:precession_comparison}
\end{figure}
This agreement is visualised in the upper panel of figure \ref{fig:precession_comparison} which compares $\omega^2$ for the Doppler shifted bending wave theory (dashed lines) and the ring model (solid lines) as $\omega_a$ is varied whilst $\epsilon=0.01$ is held fixed. We see that the upper frequency branch corresponding to the retrograde precession matches well for all epicyclic frequencies. However, the lower branches corresponding to the prograde precession begin to deviate noticeably when $X$ falls below $2/3$. 

\subsection{Precession of resonant tilting modes}

We have seen that when the epicyclic and vertical frequencies match, the resonant response leads to an enhanced precession of the ring. Indeed, the local resonant tilting mode frequency departs from the orbital rate as $\sim O(\epsilon)$ whilst the non-resonant case goes as $\sim O(\epsilon^2)$. This highlights the importance of resonances in driving distinct dynamics. 

Whilst we have considered the precessional behaviour of pure modes in the above analysis, it is important to understand the long term evolution for general initial conditions. It is useful here to proceed using the Lagrangian linear theory. Similar to the Eulerian analysis discussed in Section \ref{section:linear} we perturb the terms which break the midplane symmetry of the equilibrium. This leads to a simple eigenvalue problem for which the general solution is given by
\begin{align}
    \begin{pmatrix}
    J_{13} \\
    J_{31}
    \end{pmatrix} &= 
    c_1 \begin{pmatrix}
    1 \\
    1
    \end{pmatrix} e^{i \omega_1 t} +
    c_2 \begin{pmatrix}
    1 \\
    1
    \end{pmatrix} e^{-i \omega_1 t} \nonumber\\
    &+ c_3 \begin{pmatrix}
    -1 \\
    1
    \end{pmatrix} e^{i \omega_2 t}
    + c_4 \begin{pmatrix}
    -1 \\
    1
    \end{pmatrix} e^{-i \omega_2 t}
    +
    \frac{2C_{x0}}{1-\epsilon^2} \begin{pmatrix}
    1 \\
    -\epsilon
    \end{pmatrix}.
\end{align}
Here, $C_x = \epsilon C_{x0}$ relates to the perturbative constant of integration and $\omega_{1,2} = \sqrt{1\pm \epsilon}$ in accordance with resonant tilting frequencies identified in Section \ref{subsubsection:tilting_modes}. The eigenvectors show an equipartition between the $J_{13}$ shear and $J_{31}$ tilt contributions as per resonant bending wave theory. The constants $c_i$ are then constrained using the initial conditions $J_{ij}(0) = J_{ij,0}$ and $\dot{J}_{ij}(0) = \dot{J}_{ij,0}$. 

In order to extract the long timescale precessional behaviour from this local oscillatory solution, we require some way of isolating the global orientation of the ring. This is motivated by thinking about a test particle on an inclined orbit. Locally the vertical motion can be described by $z = Re(Ze^{-i\Omega t})$ where $Z$ is a complex amplitude and we have assumed a spherically symmetric potential for which $\nu = \Omega$. The magnitude of $Z$ is proportional to the orbit inclination whilst the argument is related to the longitude of the ascending node. Thus, we may express this quantity in terms of the classic complex tilt variable $W = l_x+i l_y$ by means of $Z = -r_0 W$. The evolution of $Z$ then tracks the global evolution of the horizontal angular momentum vector. \cite{Ogilvie2013} construct the local analogue of the horizontal angular momentum to be 
\begin{equation}
    -r_0 \dot{z} \mathbf{e}_y - r_0 \Omega_0 z \mathbf{e}_x
\end{equation}
where $\mathbf{e}_x$ and $\mathbf{e}_y$ are the radial and azimuthal unit vectors in the local model. Upon normalising against the total angular velocity for a circular orbit, $r_0^2\Omega$, and multiplying by $\exp{(i\Omega t)}$ to counteract the local frame rotation, we find 
\begin{equation}
    Z = \left(z+\frac{i \dot{z}}{\Omega}\right)e^{i\Omega t}.
\end{equation}
In the absence of any torques which drive the evolution of this orbit, angular momentum is conserved and $Z$ is a constant -- essentially describing the fixed amplitude and phase of a harmonic oscillator in terms of its displacement and velocity at any given time. 

\begin{figure}
    \centering
    \includegraphics[width=\columnwidth, trim={0 3cm 0 3cm}, clip]{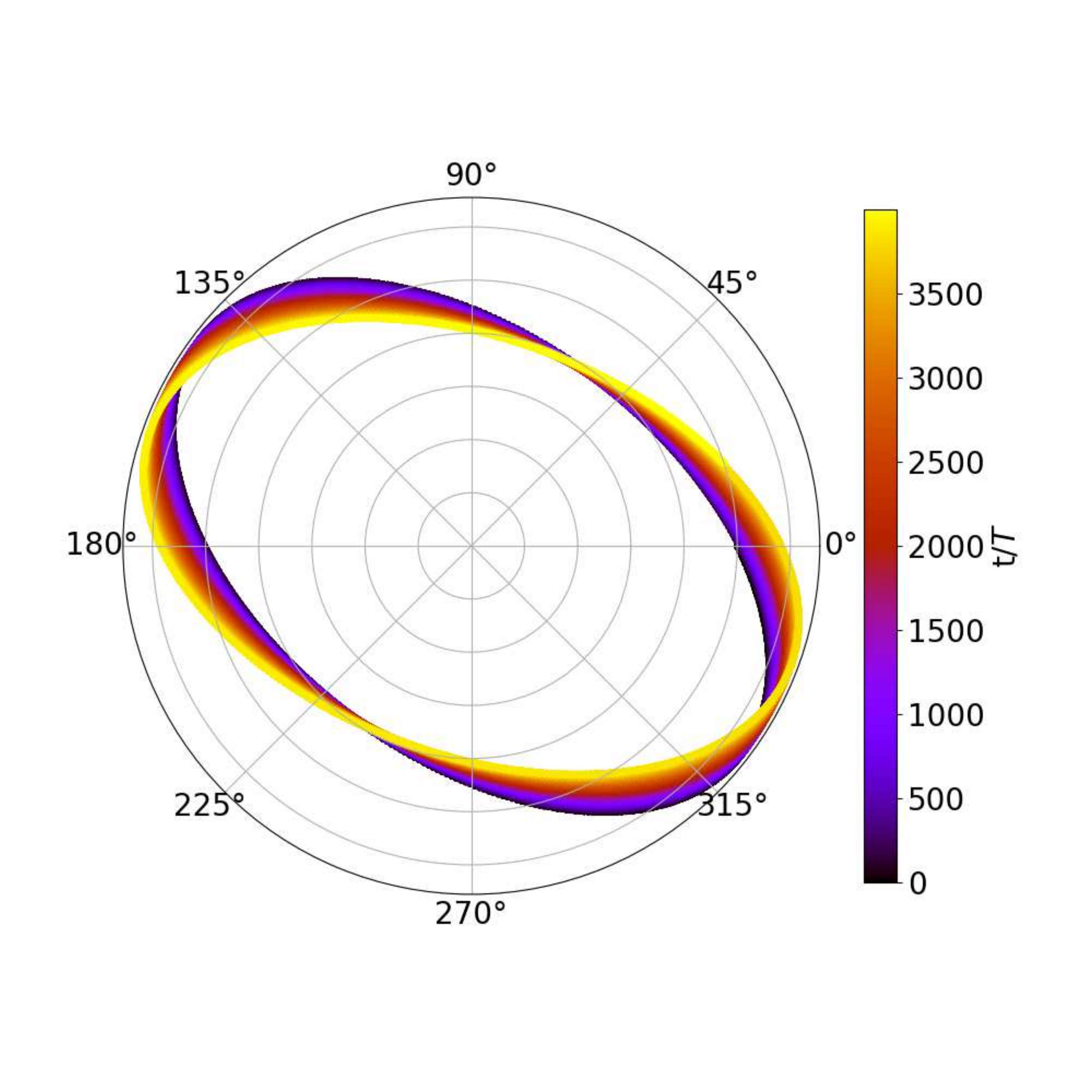}
    \caption{The reference warp amplitude $\psi_0 = dZ/dx_0$ is plotted for an arbitrary linear tilting initial condition for a ring with an equilibrium aspect ratio $\epsilon=0.01$. The colour bar represents the advancing time in units of the orbital period. To leading order $O(\epsilon)$, elliptical tracks are traced in the retrograde direction over 200 orbital periods. Over the longer timescale $\sim O(\epsilon^2)$ these elliptical tracks precess in the prograde direction, as indicated by the purple ellipses at earlier times rotating counter-clockwise to the yellow ellipses at later times.}
    \label{fig:Z_precession}
\end{figure}

More generally in our fluid continuum, we will regard the mass weighted average of $Z$ as our measure of the tilt orientation of the ring. Our Lagrangian reference state is symmetric about the midplane so a vertical averaging extracts the line $z_0 = 0$. Inserting our Lagrangian variables, elements on this line then have the complex amplitude
\begin{equation}
    Z = \left( J_{31}+\frac{i}{\Omega} \dot{J}_{31}\right)x_0 e^{i\Omega t}.
\end{equation}
The local measure of warp is then given by the gradient of this line
\begin{equation}
    \psi \equiv \frac{dZ}{dx} = \frac{dx_0}{dx}\frac{dZ}{dx_0} = \frac{J_{33}}{J}\left( J_{31}+\frac{i}{\Omega} \dot{J}_{31}\right) e^{i\Omega t}.
\end{equation}
The prefactor $J_{33}/J$ is simply constant in linear theory, corresponding to the width of the ring which scales the magnitude of the tilt. Let us instead focus on the evolution of the reference warp term
\begin{equation}
    \psi_0 \equiv  \frac{dZ}{dx_0} = \left( J_{31}+\frac{i}{\Omega} \dot{J}_{31}\right) e^{i\Omega t}.
\end{equation}
We now insert the general solution for $J_{31}$ and the resonant tilting mode frequencies expanded to first order $\omega_{1,2} = 1\pm\epsilon/2$. Expanding terms and gathering the leading order contributions shows that
\begin{equation}
    \begin{pmatrix}
    Re(\psi_0) \\
    Im(\psi_0)
    \end{pmatrix} 
    =
    \begin{pmatrix}
    J_{31,0} & \dot{J}_{13,0} \\
    \dot{J}_{31,0} & -J_{13,0}
    \end{pmatrix}
    \begin{pmatrix}
    \cos{(\epsilon t/2)} \\
    \sin{(\epsilon t/2)} \\
    \end{pmatrix}.
\end{equation}
This describes a linear transformation of a circular path, so in general the global warp quantity $\psi_0$ evolves on elliptical tracks. Only when the system is initialised in one of the normal modes, with equal amplitude $J_{31}$ and $J_{13}$ in perfect phase or anti-phase, do we see circles. This describes the precession of resonant tilting modes on a timescale $\sim O(\epsilon^{-1})$ for arbitrary initial conditions. An example of a general precession track in shown in Fig.~\ref{fig:Z_precession}. The angular coordinate measures the phase of the tilt and thus tracks the changing orientation of the longitude of the ascending node. The radial coordinate then measures the amplitude of the warp which varies as a consequence of the beating phenomenon between the two slightly detuned normal modes. For a ring with aspect ratio $\epsilon = 0.01$, a full precessional rotation occurs after 200 orbital timescales.
If we expand the resonant frequencies to second order in the aspect ratio, such that $\omega_{1,2} = 1\pm\epsilon/2 - \epsilon^2/8$, this introduces a long timescale prograde bias to the structure. Indeed, we can show that $\psi_0 \mapsto \psi_0 \exp[i\epsilon^2 t/8]$ such that the elliptical tracks rotate anticlockwise over time. This longer timescale is visualised in Fig.~\ref{fig:Z_precession} as the later time yellow tracks have advanced prograde to the earlier purple tracks. This hierarchy of timescales will be important in understanding the evolution of tilted rings in Keplerian systems and the resonant evolution of warped discs as will be explored further in a companion paper.

%% file: Sections/6_numerical.tex
\section{Numerical excitation of ring model modes}
\label{section:numerical}

Having analytically constructed our model and identified the linear modes, we will now demonstrate the excitation of these within a grid based code. Such torus oscillations are difficult to simulate given that the finite extent of the structure leads to zero density regions which are not well described on a fixed grid. However, here we are in fact able to find good agreement between our analytical ODE model and a full hydro solution found using the finite-volume code PLUTO.

PLUTO naturally lends itself to our local ring model as we can invoke the native shearing box module. Furthermore, the FARGO scheme is implemented, wherein the background Keplerian shear flow is subtracted and the code solves for the residual velocity components. This is particularly useful for thin rings where the underlying Keplerian shear flow can become large and would severely limit the integration time-step. 

\subsection{Setting up an equilibrium}
\begin{figure*}
    \centering
    \includegraphics[width=2\columnwidth]{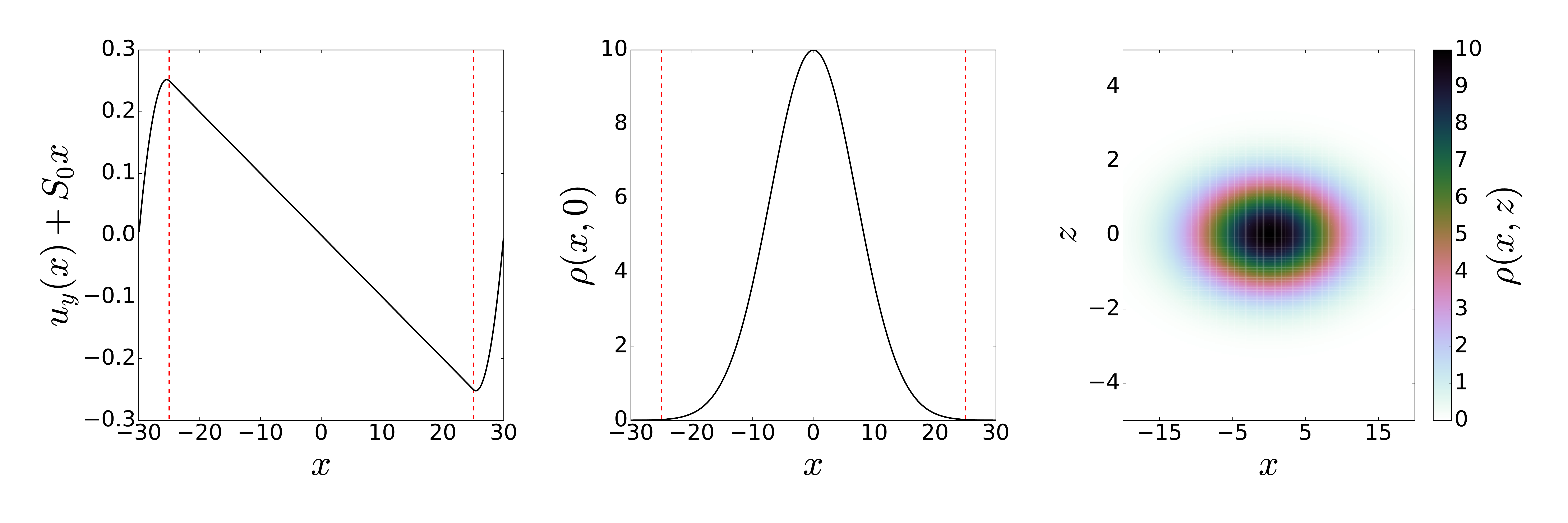}
    \caption{The modified equilibrium configuration as used in our PLUTO runs. \textit{Left Panel:} The residual azimuthal velocity, where the background Keplerian shear flow is subtracted from the actual flow field, along a central radial line $z=0$ through the ellipse. The red dashed lines indicate the position at which the zonal flow is modified to match onto the Keplerian value so the residual flow reduces to zero at the boundaries. \textit{Middle Panel:} The density profile through the centre of the ellipse. This reduces to a flat line with zero radial gradient at the boundaries. \textit{Right Panel:} A cross-sectional slice in $x-z$ visualising the elliptical equilibrium. Note that the x and z axes are differently scaled so don't visually capture the small aspect ratio.}
    \label{fig:numerical_equilibrium}
\end{figure*}

In order to avoid problems associated with zero density regions, we set up an isothermal equilibrium ring with a Gaussian density profile. This is as valid as any other choice since we found that the ring model modes are insensitive to the details of the spatial pressure and density structure set by $\tilde{p}$ and $\tilde{\rho}$. Note however, that the dynamical evolution takes place under an ideal adiabatic energy equation with $\gamma = 5/3$, which does affect the linear breathing modes (see equation \eqref{linear_eq:breathing_modes_dispersion}).

We will investigate the resonant regime for which the central potential sets $\kappa=\nu=\Omega$ and $S = (3/2)\Omega$.  We must then carefully set up an equilibrium compatible with the shearing periodic boundary conditions. This requires that the zonal flow reduces to the Keplerian value at the boundaries. Thus the equilibrium constructed in Section \ref{subsection:eulerian:simple_equilibrium} is now modified according to the piece-wise velocity field,
\begin{equation}
    \label{numerical_eq:piecewise_velocity}
    u_y =
    \begin{cases}
    -Ax & |x|<l_0 \\
    -Ax+(A-S)\frac{L_x}{2}\left( \frac{x-l_0}{L_x/2-l_0}\right)^2 &  x \ge l_0 \\
    -Ax-(A-S)\frac{L_x}{2}\left( \frac{x-l_0}{L_x/2-l_0}\right)^2 & x \le -l_0
  \end{cases}
\end{equation}
where $L_x$ is the total width of the numerical domain in $x$ and $l_0$ is an arbitrary position which delimits the transition between the linear zonal flow and Keplerian flow. This quadratic modification matches onto the Keplerian flow at the boundary. Although this breaks the linear flow field assumption used to construct the ring model, the dynamics of the modes should be unaffected provided this correction is switched on at suitably large $l_0$ where the density is negligible. Modification of the flow field is accompanied by a change in the radial profiles of density and pressure, as determined by geostrophic balance. For an isothermal equilibrium with $p = c_s^2 \rho$, where $c_s^2$ is the isothermal sound speed, we have
\begin{equation}
    \partial_x \ln{\rho} = \frac{2\Omega(u_y+S x)}{c_s^2}.
\end{equation}
The density only exhibits a slight deviation from the Gaussian in the tails such that
\begin{equation}
    \label{numerical_eq:piecewise_density}
    \rho = \rho_z(z)
    \begin{cases}
    \exp\left[-\frac{\Omega}{c_s^2}(A-S)x^2 \right] & |x|<l_0 \\
    \exp\left[-\frac{\Omega}{c_s^2}(A-S)\left(x^2-\frac{L_x}{3}\frac{(x-l_0)^3}{(L_x/2-l_0)^2}\right) \right] &  x \ge l_0 \\
    \exp\left[-\frac{\Omega}{c_s^2}(A-S)\left(x^2+\frac{L_x}{3}\frac{(x-l_0)^3}{(L_x/2-l_0)^2}\right) \right] &  x \le -l_0,
  \end{cases}
\end{equation}
where $\rho_z(z) = \rho_c\exp[-\Omega^2 z^2/2c_s^2]$ is set by the vertical hydrostatic balance and $\rho_c$ is the characteristic density value at the centre of the ring. The particular equilibrium used in our experiment is visualised in figure \ref{fig:numerical_equilibrium}. This corresponds to a disc with aspect ratio $\epsilon = 0.14$, requiring a flow coefficient $A=1.51$. For ideal hydrodynamics we are also free to choose our non-dimensional units such that $c_s=\Omega=1$. The domain is taken to have dimensions $x\in [-30,30]$ and $z\in[-5,5]$ such that $L_x=60$. The piece-wise delimiter $l_0$ is taken to be $25$, so the density has dropped off sufficiently, as shown by the red dashed lines in figure \ref{fig:numerical_equilibrium}.

\subsection{Exciting linear modes}
This equilibrium is stable when implemented in PLUTO and acts as a suitable background which can then be perturbed. We would like to excite both tilting and breathing modes and compare with the solution obtained by integrating our ring model ODEs derived in Section \ref{section:eulerian}. To this end we employ an implicit Runge-Kutta Radau solver and numerically integrate the corresponding initial value problem. We will consider two different perturbations. Taking the initial value of $A_{31}=0.01$ excites tilting mode motions in the ring as the mid-plane symmetry is broken. Meanwhile, taking $A_{33}=0.01$ initialises the ring with vertical velocity corresponding to a breathing mode.

\begin{figure*}
    \centering
    \includegraphics[width=2\columnwidth]{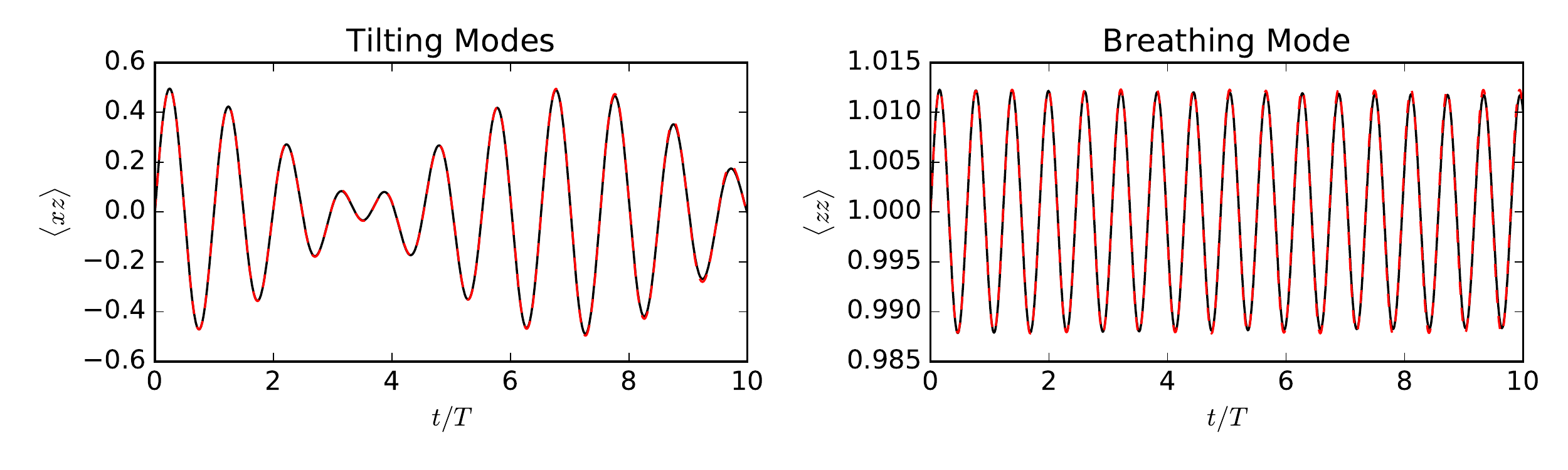}
    \caption{Comparison between the PLUTO and ODE runs for capturing the linear modes. The covariance diagnostics are numerically integrated for the PLUTO runs and shown as solid black lines whilst the analytical ODE results are plotted as dashed red lines.  \textit{Left Panel:} The tilting modes are excited with the initial perturbation $A_{31} = 0.01$ and tracked using the covariance $\langle xz \rangle$. \textit{Right Panel:} The breathing mode is excited with the perturbation $A_{33} = 0.01$ and tracked using the covariance $\langle zz \rangle$.}
    \label{fig:numerical_covariance}
\end{figure*}

In order to compare the subsequent evolution of the ring between the PLUTO and ODE schemes, we require some diagnostic which is sensitive to the small oscillations. We will examine the mass weighted covariance moments which are defined by
\begin{equation}
    \langle x_i x_j\rangle = \frac{1}{M}\int \rho x_i x_j dx dz
\end{equation}
and characterise the shape of the ellipse. Indeed, one can show that this measure is intrinsically related to the shape matrix components as follows. Consider $\langle xx \rangle$ and convert to Lagrangian coordinates:
\begin{equation}
    \langle xx \rangle = J_{11}^2 \langle x_0 x_0\rangle + J_{13}^2 \langle z_0 z_0\rangle = L^2(J_{11}^2+J_{13}^2).
\end{equation}
Likewise for $\langle xz \rangle$ and $\langle zz \rangle$
\begin{align}
    \langle xz \rangle &= L^2(J_{11}J_{31}+J_{13}J_{33}), \\
    \langle zz \rangle &= L^2(J_{31}^2+J_{33}^2).
\end{align}
Now the Lagrangian coordinates can be related to the shape matrix using expression \eqref{lagrangian_eq:J_S_relation}. Solving this system of matrix equations gives
\begin{align}
    S_{11} &=  \frac{1}{J^2 L^2}(J_{31}^2+J_{33}^2), \\
    S_{13} &= -\frac{1}{J^2 L^2}(J_{11}J_{31}+J_{13}J_{33}), \\
    S_{33} &=  \frac{1}{J^2 L^2}(J_{11}^2+J_{13}^2)
\end{align}
whilst taking the determinant gives 
\begin{equation}
    \frac{1}{J^2 L^4} =\det(S_{ij}) = S_{11}S_{33}-S_{13}^2 \equiv |\mathbfss{S}|.
\end{equation}
Combining these relations allows us to write
\begin{equation}
    \label{numerical_eq: covariance_shape_matrix}
    \langle xx \rangle =  \frac{S_{33}}{|\mathbfss{S}|}, \quad
    \langle xz \rangle = -\frac{S_{13}}{|\mathbfss{S}|} \quad \text{and} \quad
    \langle zz \rangle =  \frac{S_{11}}{|\mathbfss{S}|}.
\end{equation}
%
%
These are plotted as the dashed red lines in Fig.~\ref{fig:numerical_covariance}. Meanwhile the numerical covariance results, plotted as solid black lines, are calculated by numerical integration of the mass distribution across the grid. In the left hand panel we see the diagonal covariance moment which tracks the dominant motion for the tilting mode run. We see almost perfect overlap between the ODE and PLUTO runs over the course of 10 orbital periods. The beating phenomenon seen indicates that both tilting modes are excited with the beating interval indicative of the frequency difference between the modes. Indeed, in this near resonant case with a relatively thin ring equation \eqref{linear_eq:tilting_modes_resonant} predicts $\Delta\omega \sim \epsilon$ which corresponds to a beating interval of $t_b/T = 1/\epsilon$. For our chosen ring this corresponds to $t_b/T \approx 7.1$ which agrees well with the pattern observed. Similarly, in the right hand panel we examine the vertical covariance moment and again observe good agreement between PLUTO and the ODE comparison. In this case, the kick to $A_{33}$ only excites the vertical breathing mode and a single frequency is present. 
%

%% file: Sections/7_discussion.tex
\section{Connection with previous theory}
\label{section:discussion}
In this work we have developed a novel ring model for investigating hydrodynamical phenomena. The development of this model, in part, has been inspired by the historical treatment of equilibrium fluid figures. Indeed, the elliptical cylinders constructed in this paper are reminiscent of the classical work on homogeneous ellipsoids, as summarised by \cite{Chandrasekhar1969}. Whilst this classical work incorporates self-gravity, which is easier to do in the case of a homogeneous, incompressible fluid, we have so far neglected this effect. Many astrophysical discs are non-self-gravitating, such that their dynamics is dominated by the tidal gravitational potential and the fluid pressure, as in our model. By comparing the pressure term and destabilising self-gravity term in the 2D dispersion relation for linear density waves we find that the mass ratio of the ring to the central star should satisfy $M_{ring}/M_* << (H/R)^2$. However, young massive protoplanetary discs are susceptible to the gravitational instability and in future work we will extend our model to include these neglected terms. Indeed, the modular nature of the Lagrangian formalism allows for additional physics to easily be incorporated. 

Roche and Darwin ellipsoids which include tidal effects have since been explored in a host of astrophysical applications. \cite{CarterLuminet1983} developed an affine model for a star being tidally disrupted within the tidal radius of a black hole. Akin to our Jacobian matrix, they restrict the allowed dynamical degrees of freedom to components of a deformation matrix which then allows for a Lagrangian formalism. This affine model crucially permits compressible deformations of the star, which becomes squashed in a `pancake' phase as the star passes through pericentre. Subsequently, the dynamical model of viscous tidal disruption developed by \cite{Sridhar1992} also bears formal similarity to our oscillating ring. They examined an ellipsoidal planetisimal described by a time dependent shape matrix in a locally expanded tidal potential. By assuming a flow field linear in the Cartesian coordinates, they reduce the fluid equations of motion and Poisson's equation to a set of coupled differential equations. Whilst their study assumes an incompressible, uniform density ellipsoid, the work of \cite{Goodman1987} models blobby fluid `\textit{planets}' as polytropic ellipsoids and finds the equilibrium structure corresponds to linear flow fields. In fact, their paper is addressing the nonlinear outcome of the non-axisymmetric breakup of tori due to the Papaloizou-Pringle instability \citep{PapaloizouPringle1983} which may lead one to question the validity of our axisymmetric ring model. However, this instability is suppressed for steep angular momentum gradients \citep{PapaloizouPringle1985} for which our torus has a small aspect ratio. Thus the thin ring regime, in which we are most interested for the application to warped discs, is resistant to such breakups. 

More recently, \cite{Ogilvie2018} has developed an an affine model of the dynamics of thin discs, which incorporates vertical degrees of freedom wherein the evolution is described at each radial location by the time-dependent affine transformation of a fluid column. We can show that our ring model is a special case of this more general theory. Specifically the model must make the assumption that the deformation gradient tensor is constant in each fluid column. However, in our model this approximation is exact as the contraction and expansion of the ring is homogeneous with $J$ having no spatial dependence. A detailed comparison between the our model and this previous work is presented in Appendix \ref{appendices:affine}.

Our analysis of linear modes in Section \ref{section:linear} agrees with the lowest order global modes of slender tori identified by the analysis of \cite{Blaes2006} in the non-relativistic limit. The so called \textit{X} and inertial modes correspond to our incompressible tilt and shear modes whilst the $+$ and breathing modes correspond to our radial and vertical compressive modes respectively. A more general perturbation of our equilibrium fluid ring yields an eigenvalue problem which permits higher order polynomial eigenfunctions. These are of course excluded by our linear flow field assumption.

%% file: Sections/8_conclusion.tex
\section{Conclusion}
\label{section:conclusion}

In this paper we have presented a novel analytical framework for exploring a range of hydrodynamic phenomenon, with particular application to torus oscillations and warped disc theory. We have constructed a local ring model within a shearing sheet approximation, whereby variation of the aspect ratio can probe both the thick torus and thin ring regime. This is introduced from an Eulerian perspective and then using a Lagrangian formalism. We have calculated the linear modes and drawn a close correspondence with linear bending wave theory. The tilting ring within the local model can be interpreted as a precessing inclined disc as seen from a global view, with both prograde and retrograde motion depending on whether the tilt and shear are in phase or anti-phased. The existence of these modes is also identified within a grid based numerical simulation, which highlights the difficulties of treating boundary conditions. Instead our model allows for a dynamical evolution via the comparatively simple task of solving a set of coupled ordinary differential equations. In a future paper we will exemplify the key features of this model by exploring the extremely nonlinear dynamics of a thin torus. This will reveal a strong mode coupling between warping motions and vertical bouncing of the disc. This phenomenon requires further attention in misaligned disc systems with regards to future numerical experiments and observational implications. 

%% file: Sections/acknowledgements.tex
\section*{Acknowledgements}
The authors would like to thank the anonymous reviewer for their helpful comments and suggestions. This research was supported by an STFC studentship and STFC grants ST/P000673/1 and ST/T00049X/1.

%% file: Appendices/affine.tex
\section{Comparison with affine model}
\label{appendices:affine}

Our ring model presents a special case of the affine model for warped discs as developed by \cite{Ogilvie2018}, hereafter OL18. Indeed, our Lagrangian formalism can be derived from this theory upon introducing some simplifying assumptions about the reference state and the nature of the linear Jacobian mapping. In this Section we will compare the two models and identify the necessary restrictions to obtain equations \eqref{lagrangian_eq:J11_ode} -- \eqref{lagrangian_eq:J33_ode}.

\subsection{Affine mapping}

\begin{figure}
    \centering
    \includegraphics[width=\columnwidth]{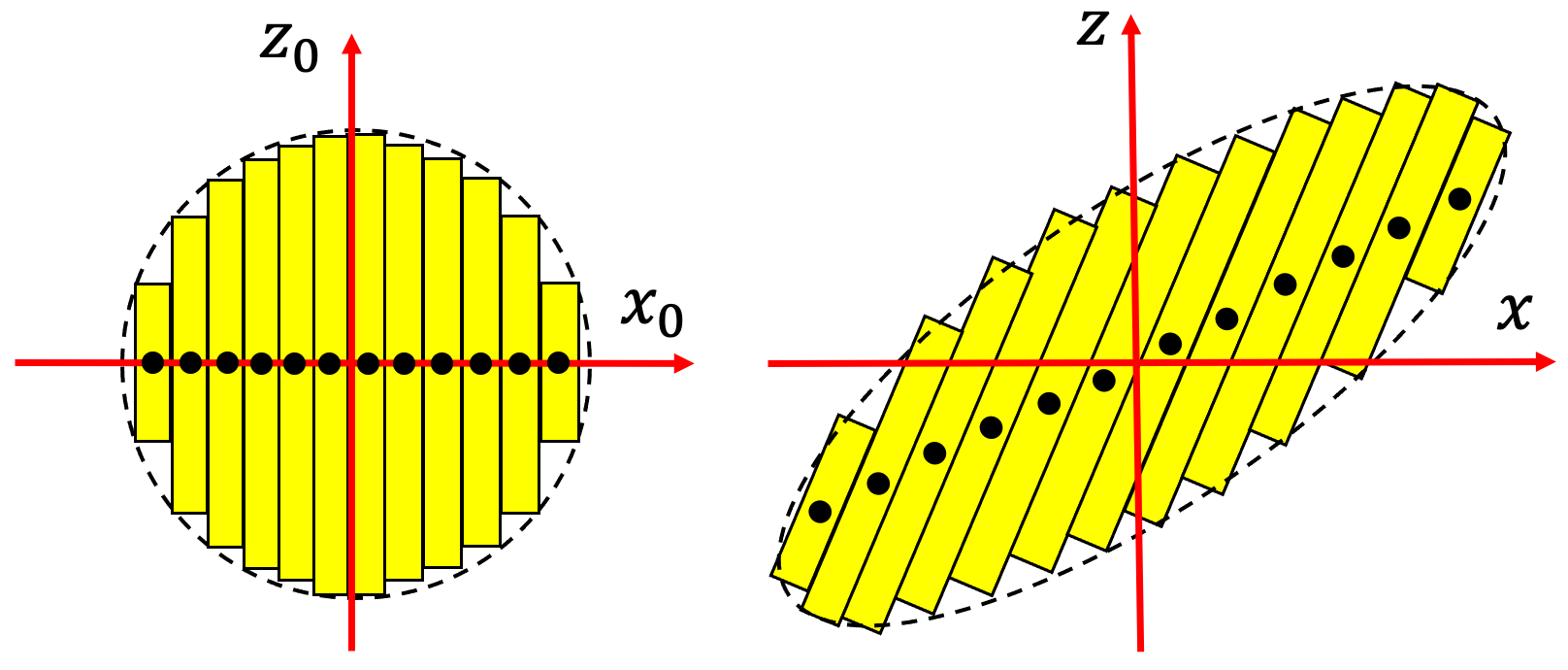}
    \caption{\textit{Left:} The reference state consists of vertical fluid columns with the height chosen to represent circular density contours. The centre of these columns is denoted by black dots lying on the midplane. \textit{Right:} Under the linear Jacobian transformation, each of these columns undergoes a translation which stretches and tilts the midplane line, followed by a rotation and stretching of the columns. This results in generally elliptical density contours.}
    \label{fig:affine:affine_cartoon}
\end{figure}

OL18 uses a Lagrangian framework to construct a model for warped discs. This involves mapping material points from a reference state to a dynamical state, under the action of a time dependent transformation, akin to the procedure in Section \ref{section:lagrangian}. Whereas we restrict attention to linear transformations, OL18 permits more general affine mappings. In this framework the reference state, described by coordinates $(x_0,z_0)$, consists of vertical fluid columns as visualised in Fig.~\ref{fig:affine:affine_cartoon}. Each fluid column is in hydrostatic equilibrium and centred on $z_0 = 0$. The distance along each column is parameterised by the scaled coordinate
\begin{equation}
    \zeta = \frac{z_0}{H_0(x_0)},
\end{equation}
where $H_0(x_0)$ is the scale-height for the fluid column labelled by the coordinate $x_0$. The affine mapping then consists of a translation, followed by stretching and rotation of the columns. This is described by 
\begin{equation}
    \mathbf{x} = \mathbf{X}(x_0, t)+\mathbf{H}(x_0, t)\zeta,
\end{equation}
in line with equation (19) in OL18. Here $\mathbf{X} = (X,Y,Z)$ denotes the position vector of the column centres in the dynamical state and $\mathbf{H} = (H_x,H_y,H_z)$ is a vector scaling and rotating each fluid column. The resulting velocity is then given by 
\begin{equation}
    \mathbf{u} = \mathbf{v}+\mathbf{w}\zeta,
\end{equation}
where  $\mathbf{v} = D\mathbf{X}/Dt$ and $\mathbf{w} = D\mathbf{H}/Dt$. Thus each column has six degrees of freedom $(\mathbf{X},\mathbf{H})$, which can be identified with our six evolutionary equations for the Jacobian coordinates. In our model, we focus on the dynamics of the largest-scale motions, which correspond to purely linear transformations of the reference state under the Jacobian mapping \eqref{lagrangian_eq: jacobian_matrix}. By assuming linear transformations, we ensure that the deformations of the columns are homogeneous and that  they cannot clash by tilting into one another. The midplane line through the centre of each column is then mapped to a straight line through the origin as described by 
\begin{equation}
    \label{eq:affine:X_Jij}
    \mathbf{X} = (J_{11},J_{21},J_{31})x_0+(0,1,0)y_0.
\end{equation}
The mapping of each column for varying $z_0$ at fixed $x_0$ then gives the correspondence with the remaining Jacobian coordinates
\begin{equation}
    \label{eq:affine:H_Jij}
    \mathbf{H} = H_0(J_{13},J_{23},J_{33}).
\end{equation}
%
\subsection{Density and pressure structure}
%
Our assumptions concerning the elliptical ring profiles will also impose conditions on the nature of the pressure and density structure entering the model of OL18. Indeed, our reference state consists of circular pressure and density contours, so the fluid column properties must be chosen to reproduce this. It should also be noted that the framework developed by OL18 considers the reference state to be in vertical hydrostatic equilibrium. We see that this is compatible with equation \eqref{lagrangian_eq: p_rho_R_relation} for the choice $\hat{T}_0 = \hat{p}_0/ \hat{\rho}_0 = \nu^2 L^2$ in which case
\begin{equation}
    \label{eq:affine:hydrostatic_condition}
    \frac{\partial p_0}{\partial z_0} =
    \hat{p}_0 \frac{d \Tilde{p}}{d R}\frac{\partial R}{z_0} = -\nu^2 L^2 \hat{\rho}_0 R \Tilde{\rho} \frac{z_0}{L^2 R} = -\rho_0 \nu^2 z_0.
\end{equation}
Furthermore, the final dynamical equations presented by OL18 are written in terms of the vertically integrated quantities
\begin{equation}
    \Sigma_0(x_0) = \int \rho_0 dz_0 , \quad P_0(x_0) = \int p_0 dz_0.
\end{equation}
When combined with equation \eqref{lagrangian_eq: p_rho_R_relation} this implies the related condition 
\begin{equation}
    \label{eq:affine:P0_Sigma0_relation}
    \frac{d P_0}{dx_0} = -\nu^2 \Sigma_0 x_0.
\end{equation}
By vertically integrating the exemplar density and pressure relationships proposed for our ring model (e.g. equation \eqref{eulerian_eq: polytropic}) we can obtain the homogeneous, polytropic and isothermal distributions for $\Sigma_0$ and $P_0$ which will automatically satisfy conditions \eqref{eq:affine:hydrostatic_condition} and \eqref{eq:affine:P0_Sigma0_relation}.
 
\subsection{Equations of motion}

With the correspondence between model quantities made, all that remains is to compute the equations of motion. The Lagrangian presented by OL18 equation (68) is slightly modified to account for the local shearing sheet perspective as used in the ring model. Indeed, as per our Lagrangian integral \eqref{lagrangian_eq:lagrangian_form}, we must include the Coriolis terms which arise from expanding about an orbiting reference frame. The Lagrangian density then becomes
\begin{multline}
    \mathcal{L} = \Big[\Big. \frac{1}{2}(v^2+w^2)+2\Omega(v_y X+w_y H_x)+\Omega S (X^2+H_x^2) \\
    -\frac{1}{2}\nu^2(z^2+H_z^2)-\frac{J^{-(\gamma-1)}P_0}{(\gamma-1)\Sigma_0}\Big.\Big]\Sigma_0
\end{multline}
Then the dynamics is governed by the Euler-Lagrange field equations 
\begin{equation}
    \frac{\partial \mathcal{L} }{\partial X_i} - \frac{D}{Dt}\frac{\partial \mathcal{L} }{\partial v_i}-\frac{\partial}{\partial x_0}\frac{\partial \mathcal{L} }{\partial (\partial X_i/\partial x_0)} = 0,
\end{equation}
\begin{equation}
    \frac{\partial \mathcal{L} }{\partial H_i} - \frac{D}{Dt}\frac{\partial \mathcal{L} }{\partial w_i}-\frac{\partial}{\partial x_0}\frac{\partial \mathcal{L} }{\partial (\partial H_i/\partial x_0)} = 0,
\end{equation}
which give the six equations of motion
\begin{equation}
    \label{eq:affine:X_eqn}
    \frac{D^2 X}{D t^2} = 2\Omega S X+2\Omega v_y-\frac{1}{\Sigma}\frac{\partial}{\partial X}\left(P+\frac{P H_x}{H}\frac{\partial Z}{\partial X}\right),
\end{equation}
\begin{equation}
    \label{eq:affine:Y_eqn}
    \frac{D^2 Y}{D t^2} = -2\Omega v_x,
\end{equation}
\begin{equation}
    \label{eq:affine:Z_eqn}
    \frac{D^2 Z}{D t^2} = -\nu^2 Z+\frac{1}{\Sigma}\frac{\partial }{\partial X}\left(\frac{P H_x}{H}\right),
\end{equation}
\begin{equation}
    \label{eq:affine:H_x_eqn}
    \frac{D^2 H_x}{D t^2} = 2 \Omega S H_x+2\Omega w_y-\frac{P}{\Sigma H}\frac{\partial Z}{\partial X},
\end{equation}
\begin{equation}
    \label{eq:affine:H_y_eqn}
    \frac{D^2 H_y}{D t^2} = -2 \Omega w_x,
\end{equation}
\begin{equation}
    \label{eq:affine:H_z_eqn}
    \frac{D^2 H_z}{D t^2} = -\nu^2 H_z+\frac{P}{\Sigma H}.
\end{equation}
Here $H = H_z-H_x dZ/dX$ is the projected scale-height quantity equivalent to
\begin{equation}
    H^2 = \int \rho(z-Z)^2 dz \bigg/ \int \rho dz.
\end{equation}
The dynamical state density and pressure are related to the prescribed reference state quantities by means of the Jacobian determinants
\begin{equation}
    \Sigma = J_{11}^{-1}\Sigma_0, \quad P = J_{11}^{-1}J^{-(\gamma-1)}P_0.
\end{equation}
Then inserting equations $\eqref{eq:affine:X_Jij}$ -- $\eqref{eq:affine:H_Jij}$ and making use of the relation $\eqref{eq:affine:P0_Sigma0_relation}$, reduces equations \eqref{eq:affine:X_eqn} -- \eqref{eq:affine:H_z_eqn} to our ring model equations \eqref{lagrangian_eq:J11_ode} -- \eqref{lagrangian_eq:J33_ode} for the choice $\hat{T}_0 = \nu^2 L^2$.

%% file: main.bbl
\begin{thebibliography}{}
\makeatletter
\relax
\def\mn@urlcharsother{\let\do\@makeother \do\$\do\&\do\#\do\^\do\_\do\%\do\~}
\def\mn@doi{\begingroup\mn@urlcharsother \@ifnextchar [ {\mn@doi@}
  {\mn@doi@[]}}
\def\mn@doi@[#1]#2{\def\@tempa{#1}\ifx\@tempa\@empty \href
  {http://dx.doi.org/#2} {doi:#2}\else \href {http://dx.doi.org/#2} {#1}\fi
  \endgroup}
\def\mn@eprint#1#2{\mn@eprint@#1:#2::\@nil}
\def\mn@eprint@arXiv#1{\href {http://arxiv.org/abs/#1} {{\tt arXiv:#1}}}
\def\mn@eprint@dblp#1{\href {http://dblp.uni-trier.de/rec/bibtex/#1.xml}
  {dblp:#1}}
\def\mn@eprint@#1:#2:#3:#4\@nil{\def\@tempa {#1}\def\@tempb {#2}\def\@tempc
  {#3}\ifx \@tempc \@empty \let \@tempc \@tempb \let \@tempb \@tempa \fi \ifx
  \@tempb \@empty \def\@tempb {arXiv}\fi \@ifundefined
  {mn@eprint@\@tempb}{\@tempb:\@tempc}{\expandafter \expandafter \csname
  mn@eprint@\@tempb\endcsname \expandafter{\@tempc}}}

\bibitem[\protect\citeauthoryear{Abramowicz \& Klu{\'{z}}niak}{Abramowicz \&
  Klu{\'{z}}niak}{2001}]{Abramowicz2001}
Abramowicz M.~A.,  Klu{\'{z}}niak W.,  2001, \mn@doi [\aap]
  {10.1051/0004-6361:20010791}, 374, L19

\bibitem[\protect\citeauthoryear{{Abramowicz}, {Jaroszynski}  \&
  {Sikora}}{{Abramowicz} et~al.}{1978}]{Abramowicz1978}
{Abramowicz} M.,  {Jaroszynski} M.,   {Sikora} M.,  1978, \aap, \href
  {https://ui.adsabs.harvard.edu/abs/1978A&A....63..221A} {63, 221}

\bibitem[\protect\citeauthoryear{{Bardeen} \& {Petterson}}{{Bardeen} \&
  {Petterson}}{1975}]{Bardeen1975}
{Bardeen} J.~M.,  {Petterson} J.~A.,  1975, \mn@doi [\apjl] {10.1086/181711},
  \href {https://ui.adsabs.harvard.edu/abs/1975ApJ...195L..65B} {195, L65}

\bibitem[\protect\citeauthoryear{{Blaes}}{{Blaes}}{1985}]{Blaes1985}
{Blaes} O.~M.,  1985, \mn@doi [\mnras] {10.1093/mnras/216.3.553}, \href
  {https://ui.adsabs.harvard.edu/abs/1985MNRAS.216..553B} {216, 553}

\bibitem[\protect\citeauthoryear{Blaes, Arras  \& Fragile}{Blaes
  et~al.}{2006}]{Blaes2006}
Blaes O.~M.,  Arras P.,   Fragile P.~C.,  2006, \mn@doi [\mnras]
  {10.1111/j.1365-2966.2006.10370.x}, 369, 1235–1252

\bibitem[\protect\citeauthoryear{{Carroll}, {McDermott}, {Savedoff}, {van Horn}
   \& {Cabot}}{{Carroll} et~al.}{1985}]{Carroll1985}
{Carroll} B.~W.,  {McDermott} P.~N.,  {Savedoff} M.~P.,  {van Horn} H.~M.,
  {Cabot} W.,  1985, \mn@doi [\apj] {10.1086/163472}, \href
  {https://ui.adsabs.harvard.edu/abs/1985ApJ...296..529C} {296, 529}

\bibitem[\protect\citeauthoryear{{Carter} \& {Luminet}}{{Carter} \&
  {Luminet}}{1983}]{CarterLuminet1983}
{Carter} B.,  {Luminet} J.~P.,  1983, \aap, \href
  {https://ui.adsabs.harvard.edu/abs/1983A&A...121...97C} {121, 97}

\bibitem[\protect\citeauthoryear{Casassus et~al.,}{Casassus
  et~al.}{2018}]{Casassus2018}
Casassus S.,  et~al., 2018, \mn@doi [\mnras] {10.1093/mnras/sty894}, 477, 5104

\bibitem[\protect\citeauthoryear{Chandrasekhar}{Chandrasekhar}{1969}]{Chandrasekhar1969}
Chandrasekhar S.,  1969, Ellipsoidal figures of equilibrium.
Yale University Press

\bibitem[\protect\citeauthoryear{{Debes} et~al.,}{{Debes}
  et~al.}{2017}]{Debes2017}
{Debes} J.~H.,  et~al., 2017, \mn@doi [\apj] {10.3847/1538-4357/835/2/205},
  \href {https://ui.adsabs.harvard.edu/abs/2017ApJ...835..205D} {835, 205}

\bibitem[\protect\citeauthoryear{Fragile, Straub  \& Blaes}{Fragile
  et~al.}{2016}]{Fragile2016}
Fragile C.,  Straub O.,   Blaes O.,  2016, \mn@doi [\mnras]
  {10.1093/mnras/stw1428}, 461, 1356

\bibitem[\protect\citeauthoryear{Frank, King  \& Raine}{Frank
  et~al.}{2002}]{Frank2002}
Frank J.,  King A.,   Raine D.,  2002, Accretion Power in Astrophysics, 3 edn.
Cambridge University Press, \mn@doi{10.1017/CBO9781139164245}

\bibitem[\protect\citeauthoryear{Goodman, Narayan  \& Goldreich}{Goodman
  et~al.}{1987}]{Goodman1987}
Goodman J.,  Narayan R.,   Goldreich P.,  1987, \mn@doi [\mnras]
  {10.1093/mnras/225.3.695}, 225, 695

\bibitem[\protect\citeauthoryear{{Hawley}, {Gammie}  \& {Balbus}}{{Hawley}
  et~al.}{1995}]{Hawley1995}
{Hawley} J.~F.,  {Gammie} C.~F.,   {Balbus} S.~A.,  1995, \mn@doi [\apj]
  {10.1086/175311}, \href
  {https://ui.adsabs.harvard.edu/abs/1995ApJ...440..742H} {440, 742}

\bibitem[\protect\citeauthoryear{Hill}{Hill}{1878}]{Hill1878}
Hill G.~W.,  1878, American Journal of Mathematics, 1, 5

\bibitem[\protect\citeauthoryear{Kato}{Kato}{2001}]{Kato2001_2}
Kato S.,  2001, \mn@doi [Publications of the Astronomical Society of Japan]
  {10.1093/pasj/53.1.1}, 53, 1

\bibitem[\protect\citeauthoryear{Katz}{Katz}{1973}]{Katz1973}
Katz J.~I.,  1973, \mn@doi [Nature Physical Science] {10.1038/physci246087a0},
  246, 87

\bibitem[\protect\citeauthoryear{Kotze \& Charles}{Kotze \&
  Charles}{2012}]{Kotze2012}
Kotze M.~M.,  Charles P.~A.,  2012, \mn@doi [\mnras]
  {10.1111/j.1365-2966.2011.20146.x}, 420, 1575

\bibitem[\protect\citeauthoryear{Kraus et~al.,}{Kraus et~al.}{2020}]{Kraus2020}
Kraus S.,  et~al., 2020, \mn@doi [Science] {10.1126/science.aba4633}, 369, 1233

\bibitem[\protect\citeauthoryear{{Lai}}{{Lai}}{1999}]{Lai1999}
{Lai} D.,  1999, \mn@doi [\apj] {10.1086/307850}, \href
  {https://ui.adsabs.harvard.edu/abs/1999ApJ...524.1030L} {524, 1030}

\bibitem[\protect\citeauthoryear{{Lubow}}{{Lubow}}{1992}]{Lubow1992}
{Lubow} S.~H.,  1992, \mn@doi [\apj] {10.1086/171877}, \href
  {https://ui.adsabs.harvard.edu/abs/1992ApJ...398..525L} {398, 525}

\bibitem[\protect\citeauthoryear{Lubow \& Ogilvie}{Lubow \&
  Ogilvie}{2000}]{Lubow2000}
Lubow S.~H.,  Ogilvie G.~I.,  2000, \mn@doi [\apj] {10.1086/309101}, 538, 326

\bibitem[\protect\citeauthoryear{{Marino}, {Perez}  \& {Casassus}}{{Marino}
  et~al.}{2015}]{Marino2015}
{Marino} S.,  {Perez} S.,   {Casassus} S.,  2015, \mn@doi [\apjl]
  {10.1088/2041-8205/798/2/L44}, \href
  {https://ui.adsabs.harvard.edu/abs/2015ApJ...798L..44M} {798, L44}

\bibitem[\protect\citeauthoryear{{Miyoshi}, {Moran}, {Herrnstein}, {Greenhill},
  {Nakai}, {Diamond}  \& {Inoue}}{{Miyoshi} et~al.}{1995}]{Miyoshi1995}
{Miyoshi} M.,  {Moran} J.,  {Herrnstein} J.,  {Greenhill} L.,  {Nakai} N.,
  {Diamond} P.,   {Inoue} M.,  1995, \mn@doi [\nat] {10.1038/373127a0}, \href
  {https://ui.adsabs.harvard.edu/abs/1995Natur.373..127M} {373, 127}

\bibitem[\protect\citeauthoryear{Nixon \& King}{Nixon \&
  King}{2012}]{Nixon2012}
Nixon C.~J.,  King A.~R.,  2012, \mn@doi [\mnras]
  {10.1111/j.1365-2966.2011.20377.x}, 421, 1201

\bibitem[\protect\citeauthoryear{{Nowak}, {Wagoner}, {Begelman}  \&
  {Lehr}}{{Nowak} et~al.}{1997}]{Nowak1997}
{Nowak} M.~A.,  {Wagoner} R.~V.,  {Begelman} M.~C.,   {Lehr} D.~E.,  1997,
  \mn@doi [\apjl] {10.1086/310534}, \href
  {https://ui.adsabs.harvard.edu/abs/1997ApJ...477L..91N} {477, L91}

\bibitem[\protect\citeauthoryear{Ogilvie}{Ogilvie}{2018}]{Ogilvie2018}
Ogilvie G.~I.,  2018, \mn@doi [\mnras] {10.1093/mnras/sty588}, 477, 1744–1759

\bibitem[\protect\citeauthoryear{Ogilvie \& Latter}{Ogilvie \&
  Latter}{2013}]{Ogilvie2013}
Ogilvie G.~I.,  Latter H.~N.,  2013, \mn@doi [\mnras] {10.1093/mnras/stt916},
  433, 2403–2419

\bibitem[\protect\citeauthoryear{Okazaki, Kato, Fukue, Okazaki, Kato  \&
  Fukue}{Okazaki et~al.}{1987}]{Okazaki1987}
Okazaki A.~T.,  Kato S.,  Fukue J.,  Okazaki A.~T.,  Kato S.,   Fukue J.,
  1987, Publications of the Astronomical Society of Japan, 39, 457

\bibitem[\protect\citeauthoryear{Papaloizou \& Pringle}{Papaloizou \&
  Pringle}{1983}]{PapaloizouPringle1983}
Papaloizou J. C.~B.,  Pringle J.~E.,  1983, \mn@doi [\mnras]
  {10.1093/mnras/202.4.1181}, 202, 1181

\bibitem[\protect\citeauthoryear{Papaloizou \& Pringle}{Papaloizou \&
  Pringle}{1985}]{PapaloizouPringle1985}
Papaloizou J. C.~B.,  Pringle J.~E.,  1985, \mn@doi [\mnras]
  {10.1093/mnras/213.4.799}, 213, 799

\bibitem[\protect\citeauthoryear{{Papaloizou} \& {Terquem}}{{Papaloizou} \&
  {Terquem}}{1995}]{PapaloizouTerquem1995}
{Papaloizou} J. C.~B.,  {Terquem} C.,  1995, \mn@doi [\mnras]
  {10.1093/mnras/274.4.987}, \href
  {https://ui.adsabs.harvard.edu/abs/1995MNRAS.274..987P} {274, 987}

\bibitem[\protect\citeauthoryear{Rezzolla, Yoshida, Maccarone  \&
  Zanotti}{Rezzolla et~al.}{2003}]{Rezzolla2003}
Rezzolla L.,  Yoshida S.,  Maccarone T.~J.,   Zanotti O.,  2003, \mn@doi
  [\mnras] {10.1046/j.1365-8711.2003.07018.x}, 344, L37

\bibitem[\protect\citeauthoryear{Rosenfeld et~al.,}{Rosenfeld
  et~al.}{2012}]{Rosenfeld2012}
Rosenfeld K.~A.,  et~al., 2012, \mn@doi [\apj] {10.1088/0004-637x/757/2/129},
  757, 129

\bibitem[\protect\citeauthoryear{Sakai, Hanawa, Zhang, Higuchi, Ohashi, Oya  \&
  Yamamoto}{Sakai et~al.}{2019}]{Sakai2019}
Sakai N.,  Hanawa T.,  Zhang Y.,  Higuchi A.~E.,  Ohashi S.,  Oya Y.,
  Yamamoto S.,  2019, \mn@doi [Nature] {10.1038/s41586-018-0819-2}, 565, 206

\bibitem[\protect\citeauthoryear{Sridhar \& Tremaine}{Sridhar \&
  Tremaine}{1992}]{Sridhar1992}
Sridhar S.,  Tremaine S.,  1992, \mn@doi [Icarus]
  {10.1016/0019-1035(92)90193-B}, 95, 86

\bibitem[\protect\citeauthoryear{{Stella} \& {Vietri}}{{Stella} \&
  {Vietri}}{1999}]{Stella1999}
{Stella} L.,  {Vietri} M.,  1999, \mn@doi [\prl] {10.1103/PhysRevLett.82.17},
  \href {https://ui.adsabs.harvard.edu/abs/1999PhRvL..82...17S} {82, 17}

\bibitem[\protect\citeauthoryear{{de Avellar}, {Porth}, {Younsi}  \&
  {Rezzolla}}{{de Avellar} et~al.}{2018}]{DeAvellar2018}
{de Avellar} M. G.~B.,  {Porth} O.,  {Younsi} Z.,   {Rezzolla} L.,  2018,
  \mn@doi [\mnras] {10.1093/mnras/stx3071}, \href
  {https://ui.adsabs.harvard.edu/abs/2018MNRAS.474.3967D} {474, 3967}

\makeatother
\end{thebibliography}
